\documentclass[a4paper,11pt]{article}
\pdfoutput=1 

\usepackage{jheppub} 
 \usepackage{tikz}
\usetikzlibrary{hobby}
\usepackage{mathrsfs}
\usetikzlibrary{decorations.markings,intersections,patterns}
\usepackage{epsfig,amsfonts,amssymb,setspace}
\usepackage{tikz-cd}
\usepackage{relsize}
\usepackage{cancel}

\pgfkeys{/tikz/tikzit fill/.initial=0}
\pgfkeys{/tikz/tikzit draw/.initial=0}
\pgfkeys{/tikz/tikzit shape/.initial=0}
\pgfkeys{/tikz/tikzit category/.initial=0}

\pgfdeclarelayer{edgelayer}
\pgfdeclarelayer{nodelayer}
\pgfsetlayers{background,edgelayer,nodelayer,main}

\tikzstyle{none}=[inner sep=0mm]

\newcommand{\mcal}[1]{\mathcal{#1}}

\title{\centering Hyperbolic Geometry and  Closed  Bosonic String Field Theory I\\ {\large The String Vertices Via Hyperbolic Riemann Surfaces}}

 \author{Seyed Faroogh Moosavian, Roji Pius}
\affiliation{Perimeter Institute for Theoretical Physics, Waterloo, ON N2L 2Y5, Canada}
\emailAdd{sfmoosavian@perimeterinstitute.ca \qquad rpius@perimeterinstitute.ca}

\abstract{The main geometric ingredient of the closed string field theory are the string vertices, the collections of string diagrams  describing the elementary closed string  interactions, satisfying  the quantum Batalian-Vilkovisky  master equation. They  can be characterized using the Riemann surfaces endowed with the metric solving the generalized minimal area problem. However, an adequately developed theory of such Riemann surfaces is not available yet, and consequently  description of the string vertices via Riemann surfaces with the minimal area metric fails to provide  practical tools for performing calculations. We describe an alternate construction of the string vertices  satisfying the Batalian-Vilkovisky master equation using  Riemann surfaces endowed with the metric having constant curvature $-1$ all over the surface. We argue that this construction provides an approximately gauge invariant closed string field theory.}

\begin{document} 
\maketitle
\flushbottom

\section{Introduction}
\label{sec:intro}

String field theory provides a quantum field theoretic description of the dynamics of interacting strings. The perturbative expansion string field theory amplitudes agree with the string amplitudes defined using the standard formulation of string perturbation theory whenever the latter are finite \cite{Witten:1985cc,Thorn:1988hm,Zwiebach:1992ie}. They formally agree with the standard string amplitudes if the latter are infected by infrared  divergences.  The main advantage of string field theory compared to the conventional string perturbation theory  is that it allows us to use the standard quantum field theory methods for defining  S-matrix elements that are free from  infrared divergences \cite{Friedan:1985ge,Dine:1987xk,Atick:1987gy,Dine:1987gj,Pius:2013sca,Pius:2014iaa,Sen:2014pia,Sen:2015hia,Pius:2014gza,Sen:2014dqa,Sen:2015uoa,Sen:2015uaa,Sen:2015hha,deLacroix:2017lif}. Furthermore, string field theory can be used to study the formal properties of the string theory S-matrix elements such as unitarity and analyticity \cite{Pius:2016jsl,Sen:2016ubf,Sen:2016bwe,Sen:2016uzq,Pius:2018crk, deLacroix:2018tml}. Since string field theory constructs  a  Lagrangian of string theory, it has the potential to open the door towards the non-perturbative regime of string theory \cite{Schnabl:2005gv}.  However so far there is little advance in the study of  the non-perturbative features of interacting closed strings using closed string field theory \cite{Sen:2016qap,Yang:2005rx}.\par 

Closed string field theory has a complicated gauge structure. As a result, it is essential to use the sophisticated machinery of Batalian-Vilkovisky (BV) formalism \cite{Batalin:1981jr,Batalin:1984jr,Barnich:1994db,Barnich:1994mt,Henneaux:1989jq,Henneaux:1992ig,Gomis:1994he} for its quantization.   The BV formalism  introduces an anti-field for each field in the theory. The quantization is achieved by constructing the master action, a functional of both the fields and the anti-fields, which is a solution of the quantum BV master equation.  The perturbative solution of quantum BV master  equation for the closed bosoinc string field theory in string coupling  has been  constructed in \cite{Zwiebach:1992ie}.  This construction requires finding a set of string vertices which satisfy the geometrical realization of the BV master equation.  An arbitrary string vertex is a collection of string diagrams with specific number of punctures and handles which describe the elementary interactions of closed strings. The prominent feature of such a set of string vertices is that they provide a  cell decomposition of the moduli space of Riemann surfaces. Moreover all the string diagrams in a specific cell  can be associated with a unique  Feynman diagram.\par 

 Such a decomposition for the moduli space can be achieved by using Riemann surfaces endowed with  metric that solves the generalized minimal area metric problem \cite{Zwiebach:1992ie}. The generalized minimal area problem asks for the metric of least possible area under the condition that all nontrivial closed curves on the surface be longer than or equal to some fixed length, conventionally chosen to be $2\pi$. A Riemann surface endowed with  minimal area metric has closed geodesics of length $2\pi$ that foliate the surface. These geodesics form a set of  foliation bands. Foliation bands are the annuli foliated by the homotopic geodesics. The shortest distance  between the boundaries of the foliation band is defined as its height. If the surface has no finite height foliation of height bigger than $2\pi$, then the whole string diagram corresponds to an elementary interaction. Therefore, the set of all  inequivalent genus $g$ Riemann surfaces with  $n$ punctures  endowed with  minimal area and no closed curves having length less than $2\pi$ and no finite height foliation of height bigger than $2\pi$ is  defined as the string vertex $\mathcal{V}_{g,n}$. \par
 
Unfortunately,  a concrete description of the minimal area metric is available only  for genus zero Riemann surfaces. There, the minimal area metrics  always arise from the Jenkins-Strebel quadratic differentials  \cite{Strebel}. In the case of higher genus Riemann surfaces, in contrast with metrics that arise from Jenkins-Strebel quadratic differentials, where the geodesics (horizontal trajectories) intersect in zero measure sets (critical graphs), the minimal area metrics can have bands of geodesics that cross. Therefore, for higher genus, all the  minimal area metrics are not the same as the metric that arises from the Jenkins-Strebel quadratic differentials. Moreover, beyond genus zero,  a fairly concrete description is available only in terms of the structure of the foliations by geodesics that is expected to exist.  Even at genus zero level, the explicit construction of Jenkins-Strebel quadratic differentials is a daunting task \cite{Moeller:2004yy, Headrick:2018ncs, Headrick:2018dlw}. Moreover,  a rigorous  proof of the existence for such metrics is not yet available. Consequently, at present, the formulation of closed  string field theory based on  Riemann surfaces endowed with metric solving the generalized minimal area problem is not well suited for performing  computations in closed string field theory. \par

In this paper,  we describe an alternate construction of the string vertices using the Riemann surfaces with  metric having constant curvature $-1$  all over the surface. We argue that in contrast with the theory  of Riemann surfaces endowed with the minimal area metric, the theory of   Riemann surfaces endowed with hyperbolic metric  is sufficiently developed for providing a calculable formulation of the closed string field theory. Every genus-$g$ Riemann surface with  $n$ distinguished punctures subject to the constraint $2g+n\ge 3$ admits a hyperbolic metric. Such surfaces, known as hyperbolic Riemann surfaces, can be obtained by the proper discontinuous action of a Fuchsian group   on the Poincar\'e upper half-plane \cite{Abi1}. The  Fuchsian group is a subgroup of the automorphism group of the Poincare metric on the upper half-plane.  Furthermore, the theory of the moduli space of the hyperbolic Riemann surfaces  is well suited for performing integrations over the moduli space \cite{Mirzakhani:2006fta,Mirzakhani:2006eta}. \par

The string vertex that corresponds to an elementary vertex of the closed bosonic string field theory with $g$ loops and $n$ external legs can be naively defined as the set of $n$ punctured genus $g$  Riemann surfaces endowed with  a metric having constant curvature  $-1$ and having no simple closed geodesic of length less than an infinitesimal  parameter $c_*$.   The surface obtained by the plumbing fixture of surfaces belong to the naive string vertices can be associated with a unique Feynman diagram.  These naive string vertices together with the Feynman diagrams only provide an approximate cell  decomposition of the moduli space, with a slight mismatch between the adjacent cells. The size of the mismatch is shown to be of the order  $c_*^2$.  A systematic algorithm for improving the naive string vertices  perturbatively in $c_*$ is proposed. Following this algorithm, the  string vertices with leading order corrections are obtained. The leading order corrected string vertices   together with the cells associated with different Feynman diagrams  obtained by the plumbing fixture  of surfaces belong to the naive string vertices  provide a cell decomposition of the moduli space  having no mismatch up to   the order  $c_*^2$ .   These  improved string vertices can be  used to build a consistent closed bosonic string field theory by keeping the parameter $c_*$ very small. Due to the mismatch beyond the order $c_*^2$, the closed string field theory is only approximately gauge invariant. \par

This construction closely follows the construction of gluing compatible 1PI regions inside the moduli space needed for defining the off-shell amplitudes in string perturbation theory \cite{Moosavian:2017fta}. The essential difference is that the 1PI region inside the moduli space include degenerate Riemann surfaces with non-separating degenerations, unlike the string vertices which do not include any degenerate Riemann surface. Therefore, the string vertices has more boundaries compared to the gluing compatible 1PI regions, and consequently the string vertices are needed to satisfy  more stringent conditions than the gluing compatible 1PI regions.\par

Recently, the cubic vertex of heterotic string field theory has constructed by using $SL(2,\mathbb{C})$ local coordinate maps which in turn has been used to construct the one loop tadpole string vertex in heterotic string field theory \cite{Erler:2017pgf}. The cubic string vertex defined this way differ from the cubic string vertex defined by the minimal area metric. However, a similar construction of string vertices with arbitrary number of loops and punctures is not available yet. Interstingly, the construction described in \cite{Erler:2017pgf} suggests the possibility of  using the stub parameter to set up a perturbative approximation of the string vertices, which matches with the basic idea behind the construction described in this paper.\par

This paper is organized as follows.  In   section  \ref{MasterCSFT},  we review the general construction of the quantum BV master action for the closed string field theory. In section \ref{celldecomp},  we  discuss the geometrical identity satisfied by the string vertices. In section \ref{vertices}, we present a short discussion of the hyperbolic Riemann surfaces and  the  construction of the naive string vertices using them. In section \ref{consistency}, we  check the consistency of these naive string vertices and find that  together with the Feynman diagrams,  they fail to provide the exact cell decomposition of the moduli space. In the last section \ref{fillip},  we describe a systematic procedure for  correcting the naive string vertices defined using the hyperbolic Riemann surfaces and find explicitly the leading order correction to the naive string vertices. In the appendix \ref{BV},  we briefly   review the Batalian-Vilkovisky quantization procedure. \par

\section{The quantum BV master action }\label{MasterCSFT}
Let us begin by a brief review of the construction of the quantum master action of closed  string field theory following the seminal work of Zwiebach \cite{Zwiebach:1992ie}. Closed bosonic string theory is formulated in terms of a conformal field theory (CFT) defined on Riemann surface. The worldsheet  CFT consists of two sectors: the matter  and the ghost sectors. The matter sector has  central charge $(26,26)$ and  the ghost sector has  central charge   $(-26,-26)$.  The conformal dimensions of ghost fields $c(z)$ and $\bar c(\bar z)$ are $(-1,0)$ and $(0,-1)$ respectively and that of anti-ghost fields $b(z)$ and $\bar b(\bar z)$ are $(2,0)$ and $(0,2)$ respectively.  \\

\noindent\underline{\bf{String fields}}: The basic degrees of freedom in  closed string field theory  are the closed string fields. An arbitrary string field is an arbitrary vector in the Hilbert space $\mathcal{H}$ of the worldsheet CFT. It can be expressed as an arbitrary linear superposition  of the basis states $\{|\Phi_s\rangle\}$ for $\mathcal{H}$ with even Grassmanality:
\begin{equation}
|\Psi\rangle=\sum_s|\Phi_s\rangle \psi_s.
\end{equation}   
 An arbitrary target space field  $\psi_s$   is a function of the target space coordinates. It is the component of the vector $|\Psi\rangle$ along the basis vector $|\Phi_s\rangle$.  The Grassmanality of the string field $|\Psi\rangle_s$ is even and it is same as that of the CFT operator $\Psi_s$ which creates the state by acting on the vacuum.  The ghost number of a component of the string field is same as that of the corresponding first quantized state.  Target space field $\psi_s$ entering into the string field as $|\Phi_s\rangle\psi_s$ are assigned a target space ghost number defined by  
\begin{equation}
g^t(\psi_s)=2-G_s,
\end{equation}
where $G_s$ is the ghost number of the state $|\Phi_s\rangle$.  The first quantized ghost number operator $G$ is given by 
\begin{equation}\label{ghostop}
G=3+\frac{1}{2}(c_0b_0-b_0c_0)+\sum_{n=1}^{\infty}(c_{-n}b_n-b_{-n}c_n)+\frac{1}{2}(\bar c_0\bar b_0-\bar b_0\bar c_0)+\sum_{n=1}^{\infty}(\bar c_{-n}\bar b_n-\bar b_{-n}\bar c_n),
\end{equation}
 where $b_n,c_n,{\bar b_n}$ and ${\bar c_n}$ are the modes of the following mode expansions of the ghost fields
  \begin{equation}\label{modebc}
  c(z)=\sum_n\frac{c_n}{z^{n-1}}\qquad\bar c(\bar z)=\sum_n\frac{\bar c_n}{\bar z^{n-1}} \qquad    b(z)=\sum_n\frac{b_n}{z^{n+2}}\qquad\bar b(\bar z)=\sum_n\frac{\bar b_n}{\bar z^{n+2}}.
  \end{equation}
The string fields that enter into  the BV master action of  closed string field theory  are called {\it the dynamical string fields} and are required to satisfy the following  conditions: i) must be annihilated  by both  $b_0^-$ and $L_0^-$: 
\begin{equation}\label{subsidiary condition}
b_0^-|\Psi\rangle=L_0^-|\Psi\rangle=0.
\end{equation}
This is necessary to make the closed string field theory action  invariant under the local Lorentz transformations on the worldsheet.  ii)  Must satisfy the following reality condition
\begin{equation}\label{realitycond}
(|\Psi\rangle)^{\dagger}=-\langle\Psi|,
\end{equation}
where the superscript dagger denotes the Hermitian conjugation and $\langle\Psi|$ denotes the BPZ conjugate state.  $L_n$ and $\overline L_n$ denote the Virasoro generators in the left and right moving sectors of the worldsheet theory.  They are the modes of the following mode expansion of the total energy-momentum tensor 
   \begin{equation}
   T(z)=\sum_{n}\frac{L_n}{z^{n+2}}\qquad\qquad  
    \overline{T}(\overline{z})=\sum_{n}\frac{\overline{L}_n}{\overline{z}^{n+2}},
   \end{equation}
 and 
     \begin{equation}
L_0^{\pm}=L_0\pm\overline{ L}_0 \qquad \qquad  b_0^{\pm}=b_0\pm\overline{ b}_0.
   \end{equation}

 \noindent\underline{\bf{Fields and anti-fields}}: The first step in the BV formalism is  the specification of  the fields and the anti-fields in the theory. For a quick review of the BV formalism see appendix \ref{BV}.  The fields and  anti-fields are specified by splitting the dynamical string field $|\Psi\rangle$ as
\begin{equation}\label{stringfieldanti}
|\Psi\rangle=|\Psi_-\rangle+|\Psi_+\rangle,
\end{equation}
where $|\Psi_-\rangle$ contains all the fields and $|\Psi_+\rangle$ contains all the anti-fields. Both $|\Psi_-\rangle$ and $|\Psi_+\rangle$ are annihilated by $b_0^-$ and $L_0^-$. They have the following decomposition 
\begin{align}\label{psipmdecomp}
|\Psi_-\rangle&=\sideset{}{'}\sum_{G(\Phi_s)\leq 2}|\Phi_s\rangle \psi^s\nonumber\\
|\Psi_+\rangle&=\sideset{}{'}\sum_{G(\Phi_s)\leq 2}|\tilde \Phi_s\rangle \psi^*_s
\end{align}
where  $|\tilde \Phi_s\rangle=b_0^-|\Phi^c_s\rangle$, such that  $\langle\Phi_r^c|\Phi_s\rangle=\delta_{rs}$. The state $\langle\Phi_r^c|$ is the conjugate state of $|\Phi_r\rangle$.  The sum in (\ref{psipmdecomp}) extends over the basis states $|\Phi_s\rangle$ with ghost number less than or equal to two. The prime over the summation sign reminds us that the sum is only over those states that are annihilated by $L_0^-$. The target space field $\psi^*_s$ is the anti-field that corresponds to the field $\psi^s$. The target space ghost number of the fields $g^t(\psi^s)$ takes  all possible non-negative   values and that of antifields $g^t(\psi^*_s)$ takes all possible negative values. The target space ghost numbers of a field and its antifield are related via the following relation (see \ref{antifield})
\begin{equation}\label{ghostnumberaf}
g^t(\psi_s^*)+g_t(\psi^s)=-1
\end{equation}
Therefore, the statistics of the antifield is opposite to that of the field, as it should be. \\

\noindent{\underline{\bf The master kinetic term}}: The kinetic term for the classical closed bosonic string theory is given by \cite{Zwiebach:1992ie}:
\begin{equation}\label{bckinetic}
S_{0,2}=g_s^{-2}\frac{1}{2}\langle \Psi|c_0^-Q|\Psi\rangle
\end{equation}
where $g_s$ denotes the closed string coupling.  The BRST operator $Q$  for the worldsheet CFT has the following expression
 \begin{equation}\label{BRSTB}
 Q=\int \frac{dz}{2\pi\mathrm{i}}c(z)\left(T_m(z)+\frac{1}{2}T_g(z)\right)+\int \frac{d\bar z}{2\pi\mathrm{i}}\bar c(\bar z)\left(\overline{T}_m(\bar z)+\frac{1}{2}\overline{T}_g(\bar z)\right),
 \end{equation}  
   where $T_m(z)$ and  $\overline{T}_m(\bar z)$ denote the stress tensors of the holomorphic and anti-holomorphic sectors of the matter CFT and $T_g(z)$ and $\overline{T}_g(\bar z)$ denote the stress tensors of the holomorphic and anti-holomorphic sectors of the ghost CFT.  The string fields appearing in classical kinetic term are allowed to have only ghost number 2.  Due to (\ref{subsidiary condition}), this action is Hermitian.  \par
   
   The master kinetic term satisfying the classical master equation is given by the same expression for classical kinetic term (\ref{bckinetic}). The only difference is that the string fields appearing in the master kinetic term can have any ghost number.  It is straightforward to check that by simply setting all the antifields to zero, we recover the classical kinetic term from the master kinetic term. \\

\noindent{\underline{\bf   String field theory interaction vertices}}:  The conventional formulation of the perturbative string theory computes the $g$ loop contribution to the scattering amplitude of  $n$ closed string states  by integrating the  string measure $\Omega^{(g,n)}_{6g-6+2n}$ over $\mathcal{M}_{g,n}$,  the moduli space  of  genus $g$ Riemann surfaces with $n$ punctures. The basic intuition behind closed string field theory  is that  perturbative expansion of any amplitude  in the closed string theory  can be constructed by joining the elementary interaction vertices in string field theory and the  propagators  using  the usual Feynman rules, just like in any quantum field theory. One identify the integration of the string measure over a set of  cylinders  as the string propagator. Then it is natural to {\it identify the integration of the string measure over the region inside the moduli space $\mathcal{M}_{g,n}$ in which one can not find any Riemann surface having regions that can be identified with the cylinders used for constructing the string propagator as the $g$ loop elementary  interaction vertex with $n$ external string states}.   Let us denote this region inside the moduli space $\mathcal{M}_{g,n}$ as $\mathcal{V}_{g,n}$.  Hence, the $g$-loop elementary interaction vertex $\{\Psi_1,\cdots,\Psi_n\}_{g}$ for $n$ closed string fields  can be defined as the integral of the off-shell string measure $ \Omega^{(g,n)}_{6g-6+2n}\left(|\Psi_1\rangle,\cdots,|\Psi_n\rangle\right)$ over the string vertex $\mathcal{V}_{g,n}$:
\begin{equation}\label{bstringvertex}
\{\Psi_1,\cdots,\Psi_n\}_{g}\equiv\int_{\mathcal{V}_{g,n}} \Omega^{(g,n)}_{6g-6+2n}\left(|\Psi_1\rangle,\cdots,|\Psi_n\rangle\right),
\end{equation}
 where $\Psi_1,\cdots,\Psi_n$ denotes the off-shell  closed string states $|\Psi_1\rangle,\cdots,|\Psi_n\rangle$. \par

The   off-shell string measure $ \Omega^{(g,n)}_{6g-6+2n}\left(|\Psi_1\rangle,\cdots,|\Psi_n\rangle\right)$ can be constructed using the vertex operators of arbitrary conformal dimension. Remember that the integrated vertex operator having conformal dimension zero represents a state satisfying classical on-shell condition. Hence,  the off-shell string measure depends on the choice of  local coordinates around the punctures on the Riemann surface. As a result,  the integration measure of an off-shell amplitude is not a  genuine differential form on the moduli space $\mathcal{M}_{g,n}$, since  $\mathcal{M}_{g,n}$ has no information about the  various  choices of local coordinates around the punctures. Instead, we must  consider $\Omega^{(g,n)}_{6g-6+2n}\left(|\Psi_1\rangle,\cdots,|\Psi_n\rangle\right)$  as a differential form defined on a section of a larger space $\mathcal{P}_{g,n}$. This space  is  defined as a fiber bundle over $\mathcal{M}_{g,n}$. The fiber direction of  the fiber bundle $\pi: \mathcal{P}_{g,n}\to \mathcal{M}_{g,n}$ contains the information about all possible choices of   local coordinates around the $n$ punctures on a genus $g$ Riemann surface. If we restrict ourselves to the dynamical string fields, satisfying (\ref{subsidiary condition}), then we can consider the differential form of our interest as a form defined on a section of the space $\widehat{\mathcal{P}}_{g,n}$. This space is smaller compared to the space $\mathcal{P}_{g,n}$. We can describe  $\widehat{\mathcal{P}}_{g,n}$ as a base space of the fiber bundle $\widehat \pi: \mathcal{P}_{g,n}\to \widehat{\mathcal{P}}_{g,n}$. The fiber direction knows about different choices of local coordinates around each of the $n$ punctures that differ by only a phase factor.\par

An arbitrary $p$-form on an arbitrary  section of $\widehat{\mathcal{P}}_{g,n}$ can be constructed as follows.  Choice of a section corresponds to a choice of local coordinates around the punctures  on $\mathcal{R}\in \mathcal{M}_{g,n}$. Therefore we only need to construct the tangent vectors of $\widehat{\mathcal{P}}_{g,n}$ that corresponds to the tangent vectors of  the moduli space. They are given by the Beltrami differentials spanning the tangent space of the moduli space of  Riemann surfaces \cite{Yoichi}.  Consider $p$ tangent vectors $V_1,\cdots,V_p$ of the section of  $\widehat{\mathcal{P}}_{g,n}$ and an operator-valued $p$-form $B_p$, whose contraction with the tangent vectors $V_1,\cdots,V_p$ is given by 
\begin{equation}\label{opevormbos}
B_p[V_1,\cdots,V_p]=b(\vec{v}_1)\cdots(\vec{v}_p),
\end{equation}
where
 \begin{equation}\label{bvgen2}
b(\vec{v}_{k})=\int d^2z\Big(b_{zz}\mu_{k\bar z}^z+b_{\bar z\bar z}\mu_{kz}^{\bar z}\Big).
\end{equation} 
 Here $\mu_k$ denotes the Beltrami differential associated with the moduli $t^{(k)}$ of the Riemann surfaces that belong to the  section  of $\widehat{\mathcal{P}}_{g,n}$.  \par
 
 A $p$-form on the section of $\widehat{\mathcal{P}}_{g,n}$ can be obtained by sandwiching  the operator valued $p$-form, $B_p$, constructed using (\ref{opevormbos}), between the surface state $\langle \mathcal{R}|$ and the state $|\Phi\rangle$ built by taking the tensor product of external off-shell states $|\Psi_i\rangle,~i=1,\cdots,n$  inserted at the punctures:
\begin{equation}\label{pformpgnbos}
\Omega_p^{(g,n)}(|\Phi\rangle)=(2\pi \mathrm{i})^{-(3g-3+n)}\langle\mathcal{R}|B_p|\Phi\rangle.
\end{equation}
 Surface state  $| \mathcal{R}\rangle$  associated with  $\mathcal{R}$ describes the state that is created on the boundaries of the discs $D_i,~i=1,\cdots,n$ by performing a functional integral over the fields of CFT on $\mathcal{R}-\sum_iD_i$.  The inner product between  $| \mathcal{R}\rangle$  and a state $|\Psi_1\rangle\otimes\cdots\otimes |\Psi_n\rangle\in\mathcal{H}^{\otimes n}$ 
\begin{equation}\label{innerprcft}
\langle\mathcal{R}|(|\Psi_1\rangle\otimes\cdots\otimes |\Psi_n\rangle),
\end{equation} 
is given by the $n$-point correlation function on $\mathcal{R}$ with the vertex operator for $|\Psi_i\rangle$ inserted at the $i^{th}$ puncture using the local coordinate system $w_i$ around that puncture. \par

The path integral representation of $ \Omega^{(g,n)}_{6g-6+2n}\left(|\Psi_1\rangle,\cdots,|\Psi_n\rangle\right)$ is given by 
\begin{align}\label{pathintrep}
& \Omega^{(g,n)}_{6g-6+2n}\left(|\Psi_1\rangle,\cdots,|\Psi_n\rangle\right)\nonumber\\
&=dm_1\cdots d\overline{m}_{3g-3+n}\int \mathcal{D}x^{\mu}\int\mathcal{D}c~\mathcal{D}\overline{c}~\mathcal{D}b~\mathcal{D}\overline{b}~e^{-I_m(x)-I_{gh}(b,c)}\prod_{j=1}^{3h-3+n}\left| \langle\mu_j|b\rangle\right|^2\prod_{i=1}^n\left[c\overline{c}~V_{i}(k_{i})\right]_{w_i},
\end{align}
where $\langle\mu|b\rangle=\int_{\mathcal{R}} d^2z ~\mu_{\overline{z}}^z~b_{zz}$ and $\left[c\overline{c}~V_{i}(k_{i})\right]_{w_i}$ denotes the vertex operator corresponds to the state $|\Psi_i\rangle$ inserted using the local coordinate $w_i$. $I_m(x)$ is the action for matter fields and $I_{gh}(b,c)$ is the actions for ghost fields.  $z$ is the global coordinate on $\mathcal{R}$ and $(m_1,\overline{m}_1,\cdots,m_{3g-3+n},\overline{m}_{3g-3+n})$ are the coordinates of the moduli space $\mathcal{M}_{g,n}$.

\noindent{\underline{\bf{Quantum BV master equation and its solution}}}: The quantum BV master action satisfy the following quantum BV master equation  
\begin{equation}\label{mastereqbcsft}
\frac{\partial _r S}{\partial \psi^s}\frac{\partial _l S}{\partial \psi^*_s}+\hbar\frac{\partial _r }{\partial \psi^s}\frac{\partial _l S}{\partial \psi^*_s}=0,
\end{equation}
 where the target space field $\psi^*_s$ is the anti-field for the field $\psi^s$.  The  perturbative solution of this equation in the closed string coupling $g_s$   is given by \cite{Zwiebach:1992ie}:
 \begin{equation}\label{cbstringfieldaction}
 S(\Psi)=g_s^{-2}\left[\frac{1}{2}\langle\Psi|c_0^-Q_B|\Psi \rangle+\sum_{g\geq 0}(\hbar g_s^2)^g\sum_{n\geq 1}\frac{g_s^n}{n!}\{\Psi^n\}_g\right].
 \end{equation} 
  where the string field interaction vertices satisfy the following identity 
 \begin{eqnarray}\label{mainidentitycbsft}
 -\frac{1}{2}\mathop{\mathop{\sum_{g_1+g_2=g}}_{n_1+n_2=n\geq 1}}_{n_1,n_2\geq 0}\sideset{}{'}\sum_s(-)^{\Phi_s}\frac{1}{n_1!n_2!}\{\Psi^{n_1},\Phi_s\}_{g_1}\{\Psi^{n_2},\tilde \Phi_s\}_{g_2}+\frac{1}{2}\sideset{}{'}\sum_s(-)^{\Phi_s}\{\Phi_s,\tilde \Phi_s,\Psi^n\}_{g-1}=0.\nonumber\\
 \end{eqnarray}
 All the string fields that enter into the action (\ref{cbstringfieldaction}) are dynamical string fields (\ref{stringfieldanti}) having arbitrary integer ghost number.  The summation is over all states in a complete basis of the Hilbert space of the world-sheet CFT which are annihilated by $L_0^-$ and $b_0^-$, so the prime sign.  The state $|\tilde \Phi_s\rangle=b_0^-|\Phi^c_s\rangle$ is such that  $\langle\Phi_r^c|\Phi_s\rangle=\delta_{rs}$ and the state $\langle\Phi_r^c|$ is the conjugate state of $|\Phi_r\rangle$.  \par

This identity (\ref{mainidentitycbsft}) imposes a stringent condition on the string vertices $\mathcal{V}_{g,n}$,  the region inside the moduli space over which we integrate  the off-shell string measure to obtain the elementary string field theory interaction vertex with $n$ dynamical string fields and $g$ loops. The precise definition of the string vertices and the geometric equation satisfied by them is discussed in detail in the next section.\par
  
 The quantum BV master action (\ref{cbstringfieldaction}) is invariant under  the master transformation  given by 
 \begin{equation}\label{masterclosedstringtransf}
 \delta_{B}|\Psi\rangle=-\sum_{g\geq 0,n\geq 0}\frac{\hbar^g g_s^{n+2g-1}}{n!}\sideset{}{'}\sum_s(-)^{\Phi_s}|\tilde \Phi_s\rangle\{\Phi_s,\Psi^{n}\}_g\cdot \mu,
 \end{equation}
 where $\mu$ is an anti-commuting parameter.  These gauge redundancies  can be fixed by specifying the anti-fields by a  relation of the form
 \begin{equation}\label{antifieldsgaugefix}
 \psi^*_s=\frac{\partial\Upsilon}{\partial \psi^s},
 \end{equation}
 where $\Upsilon$ is a fermionic functional of the fields and anti-fields. Then, the gauge fixed path integral for  closed string field theory is obtained by integrating only over fields, and substituting the gauge fixing condition (\ref{antifieldsgaugefix}) for the antifields  in the master action $S(\psi^s,\psi^*_s)$ given in (\ref{cbstringfieldaction}):
 \begin{equation}\label{gaugefixedpathintcbs}
 Z_{\Upsilon}=\int d\psi^s\mathrm{e}^{-\frac{1}{\bar h}S(\psi^s,\frac{\partial\Upsilon}{\partial \psi^s})}.
 \end{equation}
 With the help of master equation (\ref{mainidentitycbsft}), one can verify that this gauge fixed action is independent of the  choice of the gauge fermion $\Upsilon$. Therefore, the gauge invariance of the quantum BV master action is guaranteed only if the string vertices satisfy the geometric condition imposed by the quantum BV master equation (\ref{mastereqbcsft}). We shall describe the geometric condition that a consistent set of string vertices must satisfy in the next section.

\section{The cell decomposition of the moduli space }\label{celldecomp} 

 String vertex $ \mathcal{V}_{g,n}$  can be understood as a collection of genus $g$ Riemann surfaces with $n$ punctures that form a connected  region inside the compactified moduli space $\overline{\mathcal{M}}_{g,n}$. This region  $\mathcal{W}_{g,n}$ has the following properties \cite{Zwiebach:1992ie}:
\begin{itemize}

\item Surfaces that are  arbitrarily close to the degeneration are not included in it.

\item  Surfaces that belong to $\mathcal{V}_{g,n}$ are equipped with a specific choice of  local coordinates around each of its punctures. Local coordinates around the punctures are only defined up to a phase and are defined continuously over  $\mathcal{W}_{g,n}$.  
\
\item The assignment of the local coordinates around the punctures on  the Riemann surfaces that belong to a string vertex  are  independent of the labeling of the punctures. Moreover, if a Riemann surface $\mathcal{R}$ with labeled punctures is in $ \mathcal{V}_{g,n}$ then copies of $\mathcal{R}$ with all other inequivalent  labelings of the punctures  also must be included in $ \mathcal{V}_{g,n}$.  

\item If a Riemann surface belongs to the string vertex, then  its complex conjugate also must be included in the string vertex.  A complex conjugate Riemann surface of a Riemann surface $\mathcal{R}$ with coordinate $z$ can be obtained by using the anti-conformal map $z\to -\overline{z}$.

\end{itemize}
   \begin{figure}
\begin{center}
\usetikzlibrary{backgrounds}
\begin{tikzpicture}[scale=.4]

\draw[red,  very thick] (-4,1.5) to[curve through={(-4,2.2)..(-1,2.2)}] (-1,1.5);
\draw[pattern=north west lines, pattern color=gray] (-2.5,0) ellipse (2 and 2);
\draw[pattern=north west lines, pattern color=gray] (-10.5,0) ellipse (2 and 2);
\draw[pattern=north west lines, pattern color=gray] (-18.5,0) ellipse (2 and 2);
\draw[red,  very thick] (-12.5,0)--(-16.5,0);
\draw[pattern=north west lines, pattern color=gray] (-34,0) ellipse (2 and 2);
\draw  node[below ] at (-37,1) {$\mathlarger{\mathlarger{\mathlarger{\partial}}}$};
\draw  node[below ] at (-33,-4) {$\mathlarger{\mathlarger{\mathlarger{\mathcal{V}_{g,n}}}}$};
\draw  node[below ] at (-18,-4) {$\mathlarger{\mathlarger{\mathlarger{\mathcal{V}_{g_1,n_1}}}}$};
\draw  node[below ] at (-9,-4) {$\mathlarger{\mathlarger{\mathlarger{\mathcal{V}_{g_2,n_2}}}}$};
\draw  node[below ] at (-2,-4) {$\mathlarger{\mathlarger{\mathlarger{\mathcal{V}_{g-1,n+2}}}}$};
\draw  node[below ] at (-26,1.5) {${\mathlarger{=-\frac{1}{2}\mathop{\sum_{g_1,g_2}}_{g_1+g_2=g}\mathop{\sum_{n_1,n_2}}_{n_1+n_2=n}}}$};
\draw  node[below ] at (-6.5,1) {$\mathlarger{\mathlarger{\mathlarger{-~\Delta}}}$};
\end{tikzpicture}
\end{center}

\caption{The geometrical identity satisfied  by the string vertices. The red lines indicate the special plumbing fixture constructions.}
\label{Geometrical identity}
\end{figure}

 A consistent set of string vertices satisfy the following geometric identity, which can be understood as the geometric realization of the quantum BV master equation (\ref{mastereqbcsft}):   
 \begin{equation}\label{bvmastercond}
 \partial \mathcal{V}_{g,n}=-\frac{1}{2}\mathop{\sum_{g_1,g_2}}_{g_1+g_2=g}\mathop{\sum_{n_1,n_2}}_{n_1+n_2=n}\mathbf{ S}[\{ \mathcal{V}_{g_1,n_1}, \mathcal{V}_{g_2,n_2}\}]-\Delta\mathcal{V}_{g-1,n+2}.
 \end{equation} 
 $\partial  \mathcal{V}_{g,n}$ represents the collection of all Riemann surfaces which belongs to the boundary of $\mathcal{V}_{g,n}$. $\mathbf{ S}$ denotes the operation of summing over all inequivalent permutations of the external  punctures.  $\{ \mathcal{V}_{g_1,n_1},\mathcal{V}_{g_2,n_2}\}$ denotes the set of Riemann surfaces with the choice of local coordinates that can be glued at one of the puncture from each via the special plumbing fixture relation given by
\begin{equation}\label{specialplumbing}
zw=e^{i\theta},~0\leq\theta\leq2\pi,
\end{equation}
where $z$ and $w$ denote the local coordinates around the punctures that are being glued. The special plumbing fixture corresponds to the locus $|t|=1$ of the plumbing fixture relation
\begin{equation}\label{plumbing}
zw=t,\qquad t\in \mathbb{C},~0\leq |t|\leq 1.
\end{equation}
The resulting surface has genus $g=g_1+g_2$ and $n=n_1+n_2-2$. $\Delta$ denotes the operation of gluing a pair of  punctures on a Riemann surface via special plumbing fixture relation. The first term of (\ref{bvmastercond})  corresponds to the gluing of two distinct surfaces via the special plumbing fixture and the second terms corresponds to the special plumbing fixture applied to a single surface, see figure (\ref{Geometrical identity}).\par

The geometric condition (\ref{bvmastercond})  demands that the surfaces which belong to the boundary of the  string vertices should agree with the surfaces obtained by gluing surfaces that belong to appropriate string vertices using the special plumbing fixture relation  (\ref{specialplumbing}) {\it both in  their moduli parameters and in their local coordinates around the punctures}. Notice that both the right hand side and the left hand side of the geometric identity are of equal dimensionality. The boundary of the string vertex $\mathcal{V}_{g,n}$ in the left hand side is a subspace of the compactified moduli space $\overline{\mathcal{M}}_{g,n}$ with an orientation induced from the orientation of $\overline{\mathcal{M}}_{g,n}$.  The surfaces  belong to the right hand side of the geometrical identity correspond to Feynman diagrams built with one propagator in the limit when the propagator collapses. Remember that string vertices joined by the string propagator corresponds to Riemann surfaces constructed by the plumbing fixture of two non-degenerate  Riemann surfaces with plumbing parameter $t$ in the region $|t|\leq 1$.  We can therefore fix the orientation of the terms in the right-hand side of geometric identity (\ref{bvmastercond}) by considering them as the boundaries of the regions of $\overline{\mathcal{M}}_{g,n}$  obtained via plumbing fixture (\ref{plumbing}) with $|t|<1$ of the surfaces  belong to the string vertices. \par
    \begin{figure}
\begin{center}
\usetikzlibrary{backgrounds}
\begin{tikzpicture}[scale=.65]
\begin{pgfonlayer}{nodelayer}
		\node [style=none] (0) at (-2.75, -1.25) {};
		\node [style=none] (1) at (2.5, -1.25) {};
		\node [style=none] (2) at (-0.25, 2.5) {};
		\node [style=none] (3) at (-8, 4.5) {};
		\node [style=none] (4) at (-7.75, 4.75) {};
		\node [style=none] (5) at (7.5, 4.75) {};
		\node [style=none] (6) at (7.75, 4.5) {};
		\node [style=none] (7) at (-0.25, -6.75) {};
		\node [style=none] (8) at (0, -6.75) {};
		\node [style=none] (12) at (0, 0) {$\mathcal{W}_{0,4}$};
		\node [style=none] (23) at (0, 4) {$\overline{\mathcal{M}}_{0,4}$};
		\node [style=none] (24) at (3.75, 3.25) {$1$};
		\node [style=none] (25) at (3.75, 2.25) {$2$};
		\node [style=none] (26) at (4.25, 2.75) {};
		\node [style=none] (27) at (5, 2.75) {};
		\node [style=none] (28) at (5.5, 2.25) {$3$};
		\node [style=none] (29) at (5.5, 3.25) {$4$};
		\node [style=none] (30) at (8.5, 5.5) {s-channel};
		\node [style=none] (31) at (-6, 3.75) {$1$};
		\node [style=none] (32) at (-6, 2.25) {$2$};
		\node [style=none] (33) at (-5.5, 3.25) {};
		\node [style=none] (34) at (-5.5, 3.25) {};
		\node [style=none] (35) at (-5, 2.25) {$3$};
		\node [style=none] (36) at (-5, 3.75) {$4$};
		\node [style=none] (37) at (-5.5, 2.75) {};
		\node [style=none] (38) at (-8.75, 5.5) {u-channel};
		\node [style=none] (39) at (-0.25, -7.5) {t-channel};
		\node [style=none] (40) at (-1, -4.5) {$1$};
		\node [style=none] (41) at (-1, -5.5) {$2$};
		\node [style=none] (42) at (-0.5, -5) {};
		\node [style=none] (43) at (0.25, -5) {};
		\node [style=none] (44) at (0.75, -5.5) {$3$};
		\node [style=none] (45) at (0.75, -4.5) {$4$};
	\end{pgfonlayer}
	\begin{pgfonlayer}{edgelayer}
		\draw [thick] (0.center) to (2.center);
		\draw [thick] (2.center) to (1.center);
		\draw [thick] (0.center) to (1.center);
		\draw [thick, bend right] (4.center) to (5.center);
		\draw [thick, bend right=15] (6.center) to (8.center);
		\draw [thick, bend left=15] (3.center) to (7.center);
		\draw [thick] (24.center) to (26.center);
		\draw [thick] (26.center) to (27.center);
		\draw [thick] (27.center) to (29.center);
		\draw [thick] (27.center) to (28.center);
		\draw [thick] (26.center) to (25.center);
		\draw [thick] (31.center) to (33.center);
		\draw [thick] (34.center) to (36.center);
		\draw [thick] (34.center) to (37.center);
		\draw [thick] (37.center) to (32.center);
		\draw [thick] (37.center) to (35.center);
		\draw [thick] (42.center) to (43.center);
		\draw [thick] (43.center) to (44.center);
		\draw [thick] (42.center) to (41.center);
		\draw [thick] (40.center) to (43.center);
		\draw [thick] (42.center) to (45.center);
	\end{pgfonlayer}
\end{tikzpicture}
\end{center}

\caption{The cell decomposition of the compactified moduli space $\overline{\mathcal{M}}_{0,4}$ of four punctured spheres using string vertex and the plumbing fixtures of the string vertices. $\mathcal{W}_{0,4}$ is the region in  $\overline{\mathcal{M}}_{0,4}$ covered by the string diagrams which form the string vertex $\mathcal{V}_{0,4}$. }
\label{celldecomposition1}
\end{figure}
 If we assume that the string vertices $\mathcal{V}_{g,n}$ together with the Feynman diagrams constructed by the plumbing fixture of the surfaces  belong to the string vertices  provide a {\it single cover} of the compactified moduli space  $\overline{\mathcal{M}}_{g,n}$, then it is possible to show that $\mathcal{V}_{g,n}$  satisfy the geometrical condition (\ref{bvmastercond}) \cite{Zwiebach:1992ie}.  We shall briefly sketch the idea behind this claim. Let us denote the region of the moduli space covered by the plumbing fixture of $I$ pairs of punctures on  a set of surfaces  belong to the various string vertices by $V_{g,n;I}$. Then  the geometric equation (\ref{bvmastercond}) takes the following form
\begin{equation}\label{bvconditionfeynman}
\partial \mathcal{V}_{g,n}=-\partial_pV_{g,n;1}
\end{equation}
where $\partial_p$ denotes the operation that take us to the boundary obtained by propagator collapse $(|t|=1)$. Since we  assumed that the string vertices $\mathcal{V}_{g,n}$ together with the Feynman diagrams $V_{g,n;I},~I=1,\cdots,3g-3+n$  provide a {\it single cover} of the compactified moduli space  $\overline{\mathcal{M}}_{g,n}$ , we have the identity
\begin{equation}\label{singlecover}
\overline{\mathcal{M}}_{g,n}=\mathcal{V}_{g,n}\bigcup V_{g,n;1}\bigcup \cdots \bigcup V_{g,n;3g-3+n}
\end{equation}
where $3g-3+n$ is the maximum possible number of propagators. We can arrive at the geometrical condition by using (\ref{bvconditionfeynman}) and (\ref{singlecover}) together with  the fact that the boundary  $\partial\overline{\mathcal{M}}_{g,n}$ of the compactified moduli space $\overline{\mathcal{M}}_{g,n}$ vanishes. 
 \par

Therefore, {\it the string vertices,  satisfying the geometrical condition (\ref{bvmastercond}), together with the Feynman diagrams provide a cell decomposition of the moduli space. Moreover, integrating the off-shell string measure over each  cell can be interpreted as  a specific contribution to the string amplitude that is coming from a specific Feynman diagram.} \par

For example, the moduli space of sphere with four punctures can be divided into four regions:  one region that corresponds to the string vertex $\mathcal{V}_{0,4}$, and three regions corresponds to three different gluing of two three punctured spheres corresponding to s-channel, t-channel and u-channel (see the figure \ref{celldecomposition1}).

\section{The naive string  vertices using hyperbolic  metric}\label{vertices}

The foremost  difficulty in constructing string field theory is  to find a suitable cell decomposition of the moduli spaces of  Riemann surfaces. Any naive set of Feynman rules led to multiple or infinite over-counting of surfaces. Given a Riemann surface, we must  be able to associate to it a unique Feynman diagram. In principle, the string vertices satisfying the conditions listed in the section (\ref{celldecomp})  can be constructed using the Riemann surfaces endowed with the  metric solving the generalized minimal area problem \cite{Zwiebach:1992ie}. The generalized minimal area problem asks for the metric of least possible area under the condition that all nontrivial closed curves on the surface be longer than or equal to some fixed length, conventionally chosen to be $2\pi$. Unfortunately, as explained in the introduction, the description of the string vertices using surfaces endowed with minimal area metric, at present, has not developed enough to provide a calculable framework for closed string field theory. In this section, we shall discuss an alternate construction of the string vertices using Riemann surfaces endowed with a metric having constant curvature   $-1$. \par

Consider a  Riemann surface endowed with metric having constant curvature $-1$ all over the surface.  The uniformization theorem promises that every genus-$g$ Riemann surface $\mathcal{R}_{g,n}$ with $n$ distinguished punctures subject to the constraint $2g+n\ge 3$, can be obtained by the proper discontinuous action of a  Fuchsian group $\Gamma$ on the Poincar\'e upper half-plane $\mathbb{H}$ \cite{Abi1}: 
\begin{equation}
\mathcal{R}_{g,n}\simeq \frac{\mathbb{H}}{\Gamma} \label{Uniformization Theorem}
\end{equation}
The Poincar\'e upper half-plane $\mathbb{H}$ is the upper half-plane, $\mathbb{H}=\{z:~\mathrm{Im}~z>0\}$, endowed with the hyperbolic metric given by 
\begin{equation}\label{mDH}
ds^2=\frac{dzd\bar z}{(\mathrm{Im}z)^2}
\end{equation}
This metric has constant curvature $-1$  all over the upper half-plane.    A Fuchsian group $\Gamma$ is a subgroup of the automorphism group of the Poincar\'e upper half-plane, the projective special linear group $PSL(2,\mathbb{R})$. Riemann surfaces obtained this way are hyperbolic Riemann surfaces. They are endowed  with  metric having constant curvature $-1$ everywhere. \par

A puncture on a hyperbolic Riemann surface corresponds to the fixed point of the associated parabolic element of the Fuchsian group acting on the upper half-plane $\mathbb{H} $. A parabolic element associate with the puncture is an element of the group $PSL(2,\mathbb{R})$  with trace $\pm 2$. If the fixed point of the  parabolic element associated with the puncture  is at $z=\text{i}\infty$ on the upper half-plane $\mathbb{H}$, then  it  is given by \cite{Maskit3}
\begin{equation}
A_{\infty}= \left(\begin{array}{cc}1 & n \\0 & 1\end{array}\right)\qquad n\in \mathbb{Z}.
 \end{equation} 
$A_{\infty}$ generates the following transformation on $\mathbb{H}$:
\begin{equation}
z\to z+n.
\end{equation}
Then the natural local coordinate, up to a phase ambiguity, around the puncture that corresponds to a parabolic element whose fixed point is at  $z=\text{i}\infty$ on the upper half-plane $\mathbb{H}$, is given by 
\begin{equation}
w = e^{2\pi\mathrm{i}z} \label{local coordinate for the cusp at infinity}.
\end{equation}
As required, this choice of local coordinate  is invariant under the translation, $z \to z + 1$, which represents  the action of the generator of the corresponding parabolic element.  In terms of the local coordinate $w$, the metric around the puncture takes the form 
\begin{equation}\label{hyplmetric}
ds^2=\frac{dzd\bar{z}}{(\mathrm{Im}z)^2}=\left(\frac{|dw|}{|w|\ln|w|}\right)^2.
\end{equation}

 If the fixed point of the parabolic element is at $z=x$ on the upper half-plane $\mathbb{H}$, then  it  is given by \cite{Maskit3}
\begin{equation}
A_{x}= \left(\begin{array}{cc}1+mx & -mx^2 \\m & 1-mx\end{array}\right)\qquad m\in \mathbb{Z} \qquad x\in \mathbb{R}.
 \end{equation} 
It generates the following transformation on $\mathbb{H}$:
\begin{equation}
\frac{1}{z-x}\to \frac{1}{z-x}+m.
\end{equation}
Then the natural local coordinate for the puncture that corresponds to a parabolic element, whose fixed point is at infinity $z=x$ on the upper half-plane $\mathbb{H}$, is given by 
\begin{equation}
w = e^{-\frac{2\pi\text{i}}{z-x}} \label{local coordinate for the cusp at x}.
\end{equation}
This choice of local coordinate  is invariant under the translation, $\frac{1}{z-x} \to \frac{1}{z-x} + 1$, which represents  the action of the generator of the corresponding parabolic element.  In terms of the local  coordinate $w$, the metric around the puncture takes the form (\ref{hyplmetric}). Then we can define the naive string vertices using the Riemann surfaces endowed with the hyperbolic metric  as follows.\par

\noindent{\underline{\bf The naive string vertex $\mathcal{V}^0_{g,n}$}}:  {\it Consider $\mathcal{R}$, a hyperbolic Riemann surface  with $n$ punctures and $g$ handles, having no simple closed geodesics with geodesic length $l\leq c_*$.  Here $c_*$ is an arbitrary positive real number that is much less than one, $c_*\ll 1$.  Choose the local coordinates around the punctures on  $\mathcal{R}$ to be  $\widetilde w=e^{\frac{\pi^2}{c_*}}w$, where $w$ is the natural local coordinate induced from the hyperbolic metric on  $\mathcal{R}$. The set of all such inequivalent hyperbolic Riemann surfaces with the above-mentioned local coordinates around the punctures form the naive string vertex $\mathcal{V}^0_{g,n}$}. \par

 Notice that by varying  the value of $c^*$ we can vary the size of $\mathcal{V}^0_{g,n}$, i.e. volume of the region  $\mathcal{W}^0_{g,n}$ covered by $\mathcal{V}^0_{g,n}$ inside the moduli pace $\mathcal{M}_{g,n}$. Interestingly, varying $c_*$ can be understood as rescaling the local coordinates around the punctures. The  definition of $\mathcal{V}^0_{g,n}$  can also be stated in terms of the plumbing fixture construction. For this we note the following fact. A  very thin neighbourhood of a simple closed geodesic of length $c_*$ on a hyperbolic Riemann surface is isomorphic to a hyperbolic annulus obtained by endowing a hyperbolic metric on a plumbing collar having  plumbing parameter  $|t|=e^{-\frac{2\pi^2}{c_*}}$. Then the definition of $\mathcal{V}^0_{g,n}$ in terms of the plumbing fixture construction  is as follows.\par

 \noindent{\underline{\small\bf $\mathcal{V}^0_{g,n}$ and the plumbing fixture}:} {\it It is the union of all the hyperbolic Riemann surfaces having $g$ handles and $n$ punctures which can not obtained via the plumbing fixture of hyperbolic Riemann surfaces with at least  one plumbing fixture parameter having modulus less than or equal to $e^{-\frac{2\pi^2}{c_*}}$. }\par
 
 Thus the plumbing parameters of string propagators are allowed to  vary only from $0$ to $e^{-\frac{2\pi^2}{c_*}}$. Usually the plumbing parameter associated with a string propagator is allowed to vary from $0$ to $1$. Therefore it is natural to introduce another set of plumbing parameters $\widetilde{t}_i$ which vary from $0$ to $1$.  This change of parameters corresponds to  {\it choosing the local coordinates as the one that is induced from the hyperbolic metric on the surface with the scaling factor of $e^{\frac{\pi^2}{c_*}}$}. 
 
\subsection{Examples}
Let us demonstrate the explicit construction of $\mathcal{V}^0_{g,n}$ by constructing the simplest string vertices. \par

  \begin{figure}
\begin{center}
\usetikzlibrary{backgrounds}
\begin{tikzpicture}[scale=.75]
	\begin{pgfonlayer}{nodelayer}
		\node [style=none] (0) at (-9, 5) {};
		\node [style=none] (1) at (-9, 0) {};
		\node [style=none] (2) at (-7, 0) {};
		\node [style=none] (3) at (-5, 0) {};
		\node [style=none] (4) at (-5, 5) {};
		\node [style=none] (5) at (-9, -0.5) {$-1$};
		\node [style=none] (6) at (-7, -0.5) {$0$};
		\node [style=none] (7) at (-5, -0.5) {$1$};
		\node [style=none] (8) at (-11, 0) {};
		\node [style=none] (9) at (-3, 0) {};
		\node [style=none] (10) at (-7, 3) {};
		\node [style=none] (11) at (-7, 3) {$\mathcal{F}_{0,3}$};
		\node [style=none] (13) at (-7, 6) {$\infty$};
		\node [style=none] (14) at (-3, 6) {$\tau$};
		\node [style=none] (15) at (-3.25, 6.25) {};
		\node [style=none] (16) at (-3.25, 5.75) {};
		\node [style=none] (17) at (-2.75, 5.75) {};
		\node [style=none] (18) at (-5.75, 1) {};
		\node [style=none] (19) at (-5, 2) {};
		\node [style=none] (20) at (-7.25, 0.75) {};
		\node [style=none] (21) at (-5, 3.5) {};
		\node [style=none] (22) at (-8.25, 1) {};
		\node [style=none] (23) at (-5.25, 4.75) {};
		\node [style=none] (24) at (-9, 1.5) {};
		\node [style=none] (25) at (-6.25, 5) {};
		\node [style=none] (26) at (-9, 3) {};
		\node [style=none] (27) at (-7.5, 5) {};
	\end{pgfonlayer}
	\begin{pgfonlayer}{edgelayer}
		\draw [thick] (0.center) to (1.center);
		\draw [thick] (4.center) to (3.center);
		\draw [thick, bend left=90, looseness=1.75] (1.center) to (2.center);
		\draw [thick, bend left=90, looseness=1.75] (2.center) to (3.center);
		\draw (8.center) to (9.center);
		\draw (15.center) to (16.center);
		\draw (16.center) to (17.center);
		\draw (26.center) to (27.center);
		\draw (24.center) to (25.center);
		\draw (22.center) to (23.center);
		\draw (20.center) to (21.center);
		\draw (18.center) to (19.center);
	\end{pgfonlayer}\end{tikzpicture}
\end{center}

\caption{$\mathcal{F}_{0,3}$ is the fundamental domain of the modular group $\Gamma(2)$, the Fuchsian group representing the thrice punctured sphere, in the upper half plane $\mathbb{H}$. $z=0$ corresponds to the puncture at $u=1$ on the thrice punctured sphere, where $u\in \hat{\mathbb{C}}$ .  $z=\infty$ corresponds to the puncture at $u=0$ and the points $z=1$ and $z=-1$ correspond to the puncture at $u=\infty$.}
\label{funddomain3psphere}
\end{figure}

\noindent{\underline{\bf Naive string vertex $\mathcal{V}^0_{0,3}$}}: The naive string vertex $\mathcal{V}^0_{0,3}$ contain only one surface, a  thrice punctured sphere endowed with hyperbolic metric. The hyperbolic thrice punctured sphere is obtained by considering the quotient of $\mathbb{H}$ with respect to the modular group $\Gamma(2)$ generated by the transformations
\begin{equation}
z\to \frac{z}{2z+1}\qquad \qquad z\to z+2.
\end{equation}

The fundamental domain $\mathcal{F}_{0,3}$ of the modular group $\Gamma(2)$, the Fuchsian group representing the thrice punctured sphere, in the upper half plane $\mathbb{H}$ is as shown in figure \ref{funddomain3psphere}. The point $z=0$ corresponds to the puncture at $u=1$ on the thrice punctured sphere.  $z=\infty$ corresponds to the puncture at $u=0$ and the points $z=\pm 1$ correspond to the puncture at $u=\infty$. The local coordinates around the punctures are as follows
\begin{align}\label{v03localc}
w_1&=e^{\frac{\pi^2}{c_*}}e^{2\pi\text{i}z}  &z&=\infty,\nonumber\\
w_2&=e^{\frac{\pi^2}{c_*}}e^{-2\pi\text{i}/z}  &z&=0,\nonumber\\
w_3&=e^{\frac{\pi^2}{c_*}}e^{-2\pi\text{i}/(z\pm 1)}  &z&=\mp 1.
\end{align}

The naive string vertex $\mathcal{V}^0_{0,3}$ is precisely the Kleinian vertex discussed in \cite{Sonoda:1989sj}.

\noindent{\underline{\bf Naive string vertex $\mathcal{V}^0_{0,4}$}}:  The naive string vertex $\mathcal{V}^0_{0,4}$ is a collection of four punctured hyperbolic spheres. All the inequivalent four punctured hyperbolic sphere can be obtained by varying the Fenchel-Nielsen length and twits parameters  $\ell$ and $\tau$, where $\ell\in \mathbb{R}_+, \tau\in\mathbb{R}$. The Fuchsian group $\Gamma_{0,4}(\ell,\tau)$ that produces a four punctured sphere with Fenchel-Nielsen parameter $(\ell,\tau)$ can be generated using the following three elements \cite{Maskit4}:

  \begin{figure}
\begin{center}
\usetikzlibrary{backgrounds}
\begin{tikzpicture}[scale=.75]
	\begin{pgfonlayer}{nodelayer}
		\node [style=none] (0) at (3.5, 0) {};
		\node [style=none] (1) at (-9, 0) {};
		\node [style=none] (2) at (-7, 0) {};
		\node [style=none] (3) at (-5, 0) {};
		\node [style=none] (4) at (-0.5, 0) {};
		\node [style=none] (6) at (-2.75, -0.5) {$0$};
		\node [style=none] (8) at (-10.75, 0) {};
		\node [style=none] (9) at (5, 0) {};
		\node [style=none] (10) at (-7, 3) {};
		\node [style=none] (11) at (-2.75, 3.5) {$\mathcal{F}_{0,4}(\ell,\tau)$};
		\node [style=none] (14) at (4.25, 6) {$z$};
		\node [style=none] (15) at (4, 6.25) {};
		\node [style=none] (16) at (4, 5.75) {};
		\node [style=none] (17) at (4.75, 5.75) {};
		\node [style=none] (18) at (1.5, 0) {};
		\node [style=none] (19) at (3.5, 0) {};
		\node [style=none] (20) at (-7.25, 0.75) {};
		\node [style=none] (21) at (-3.25, 5.5) {};
		\node [style=none] (22) at (-8.25, 1) {};
		\node [style=none] (23) at (-4.75, 5.25) {};
		\node [style=none] (24) at (-8.75, 1.75) {};
		\node [style=none] (25) at (-6.5, 4.5) {};
		\node [style=none] (26) at (-0.5, 0) {};
		\node [style=none] (27) at (1.5, 0) {};
		\node [style=none] (28) at (-5.75, 1) {};
		\node [style=none] (29) at (-2, 5.5) {};
		\node [style=none] (30) at (-4, 1.75) {};
		\node [style=none] (31) at (-1, 5.25) {};
		\node [style=none] (32) at (-2.5, 2) {};
		\node [style=none] (33) at (0, 5) {};
		\node [style=none] (34) at (-1.5, 1.75) {};
		\node [style=none] (35) at (1, 4.5) {};
		\node [style=none] (36) at (-0.75, 1) {};
		\node [style=none] (37) at (1.75, 3.75) {};
		\node [style=none] (38) at (0.75, 1) {};
		\node [style=none] (39) at (2.5, 3) {};
		\node [style=none] (40) at (1.5, 0.5) {};
		\node [style=none] (41) at (3, 2) {};
		\node [style=none] (42) at (-2.75, 0) {};
		\node [style=none] (43) at (-2.75, 7) {};
		\node [style=none] (44) at (-2.5, 5.25) {};
		\node [style=none] (45) at (-2.5, 2.25) {};
		\node [style=none] (46) at (-7.5, 1) {};
		\node [style=none] (47) at (-6.5, 1) {};
		\node [style=none] (48) at (1, 1) {};
		\node [style=none] (49) at (2, 1) {};
		\node [style=none] (50) at (-8.75, 1.25) {};
		\node [style=none] (51) at (-8.25, 1.25) {};
		\node [style=none] (52) at (-5.5, 1) {};
		\node [style=none] (53) at (-4.75, 1) {};
		\node [style=none] (54) at (-0.75, 1) {};
		\node [style=none] (55) at (0, 1) {};
		\node [style=none] (56) at (2.75, 1.25) {};
		\node [style=none] (57) at (3.25, 1.25) {};
		\node [style=none] (58) at (-2, 4.25) {$a_4$};
		\node [style=none] (59) at (-8.25, 1.75) {$a_7$};
		\node [style=none] (60) at (-7, 1.75) {$a_2$};
		\node [style=none] (61) at (-5, 1.75) {$a_3$};
		\node [style=none] (62) at (-0.25, 1.75) {$a_6$};
		\node [style=none] (63) at (1.5, 1.75) {$a_1$};
		\node [style=none] (64) at (3, 1.75) {$a_5$};
		\node [style=none] (65) at (1.5, -0.5) {$1$};
		\node [style=none] (66) at (-0.5, -0.5) {$e^{-\ell}$};
		\node [style=none] (67) at (3.5, -0.5) {$e^{\ell}$};
		\node [style=none] (68) at (-5, -0.5) {$-e^{-\ell+2\tau}$};
		\node [style=none] (69) at (-7, -0.5) {$-e^{2\tau}$};
		\node [style=none] (70) at (-9, -0.5) {$-e^{\ell+2\tau}$};
	\end{pgfonlayer}
	\begin{pgfonlayer}{edgelayer}
		\draw [thick, bend right=90, looseness=1.50] (0.center) to (1.center);
		\draw [thick, bend right=90, looseness=1.50] (4.center) to (3.center);
		\draw [thick, bend left=90, looseness=1.75] (1.center) to (2.center);
		\draw [thick, bend left=90, looseness=1.75] (2.center) to (3.center);
		\draw [in=180, out=0] (8.center) to (9.center);
		\draw (15.center) to (16.center);
		\draw (16.center) to (17.center);
		\draw [thick, bend left=90, looseness=1.50] (26.center) to (27.center);
		\draw (24.center) to (25.center);
		\draw (22.center) to (23.center);
		\draw (20.center) to (21.center);
		\draw [thick, bend left=90, looseness=1.75] (18.center) to (19.center);
		\draw (28.center) to (29.center);
		\draw (30.center) to (31.center);
		\draw (32.center) to (33.center);
		\draw (34.center) to (35.center);
		\draw (36.center) to (37.center);
		\draw (38.center) to (39.center);
		\draw (40.center) to (41.center);
		\draw (43.center) to (42.center);
		\draw [red, <->] (44.center) to (45.center);
		\draw [red, <->, in=105, out=75, looseness=1.50] (46.center) to (47.center);
		\draw [red, <->, bend left=75, looseness=1.50] (48.center) to (49.center);
		\draw [red, <->, in=105, out=75, looseness=1.50] (50.center) to (51.center);
		\draw [red, <->, in=105, out=75, looseness=1.50] (52.center) to (53.center);
		\draw [red, <->, in=105, out=75, looseness=1.50] (54.center) to (55.center);
		\draw [red, <->, in=105, out=75, looseness=1.50] (56.center) to (57.center);
	\end{pgfonlayer}\end{tikzpicture}
\end{center}

\caption{$\mathcal{F}_{0,4}(\ell,\tau)$ is the fundamental domain of the group $\Gamma_{0,4}(\ell,\tau)$, the Fuchsian group representing the fo punctured sphere with Fenchel-Nielsen parameters $(\ell,\tau)$, in the upper half plane $\mathbb{H}$. The boundaries of $\mathcal{F}_{0,4}(\ell,\tau)$ are identified under the action of the elements $a_1,a_2,a_3,a_4,a_5,a_6$ and $a_7$}
\label{funddomain3psphere}
\end{figure}
\begin{align}\label{Gamma04lt}
a_1&=\left(\begin{array}{cc}1+\beta & -\beta \\  \beta  & 1-\beta \end{array}\right)  \nonumber\\ 
    a_2&=\left(\begin{array}{cc}\left(1-\beta\right)& -\beta e^{2\tau}\\  \beta e^{-2\tau} &  \left(1+\beta\right)\end{array}\right) \nonumber\\
     a_3&=-\left(\begin{array}{cc}(1+\beta )e^{\ell}& \beta e^{-\ell+2\tau}\\  -\beta e^{\ell-2\tau}& (1 -\beta) e^{-\ell}\end{array}\right),
\end{align}
where $\beta=-\frac{\text{cosh}\ell+1}{\text{sinh}\ell}$. The boundaries of the fundamental domain $\mathcal{F}_{0,4}(\ell,\tau)$ of $\Gamma_{0,4}(\ell,\tau)$ in $\mathbb{H}$ are identified under the action of the elements $a_1,a_2,a_3,a_4,a_5,a_6$ and $a_7$, where $a_4,a_5,a_6$ and $a_7$ are as follows
\begin{align}\label{Gamma04lt1}
a_4&=\left(\begin{array}{cc}e^{\ell} & 0 \\0 & e^{-\ell}\end{array}\right)  \nonumber\\
 a_5&=-a_1^{-1}a_4^{-1}=-\left(\begin{array}{cc}\left(1-\beta\right)e^{-\ell}& \beta e^{\ell}\\  -\beta e^{-\ell} &  \left(1+\beta\right)e^{\ell}\end{array}\right) \nonumber\\
  a_6&=-a_4^{-1}a_1^{-1}=-\left(\begin{array}{cc}\left(1-\beta\right)e^{-\ell}& \beta e^{-\ell}\\  -\beta e^{\ell} &  \left(1+\beta\right)e^{\ell}\end{array}\right) \nonumber\\
    a_7&=-\left(\begin{array}{cc}\left(1+\beta\right)e^{\ell}& \beta e^{\ell+2\tau}\\  -\beta e^{-\ell-2\tau} &  \left(1-\beta\right)e^{-\ell}\end{array}\right).
\end{align}
The transformations $a_1,a_2,a_3, a_5,a_6$ and $a_7$ are parabolic. The fixed points  of them are at $1, -e^{2\tau}, -e^{-\ell+2\tau}, e^{\ell}, e^{-\ell}$ and $-e^{\ell+2\tau}$ respectively. They represents the four punctures on the sphere. Therefore, the local coordinate around the puncture can be defined by specifying local coordinates in the neighbourhood of these fixed points. Around a fixed point at $z=x_i$ we define the local coordinates to be 
\begin{align}\label{v04localc}
w_i=e^{\frac{\pi^2}{c_*}}e^{-\frac{2\pi\text{i}}{(z-x_i)}}.
\end{align}
This defines a string diagram which belongs to the naive string vertex $\mathcal{V}^0_{0,4}$. By varying the Fenchel-Nielsen parameters $\ell$ and $\tau$ we can obtain all the string diagrams in $\mathcal{V}^0_{0,4}$. However, there is no simple domain parameterized by the coordinates $(\ell,\tau)$ that represent $\mathcal{W}^0_{0,4}$, the connected region in the moduli space $\mathcal{M}_{0,4}$ covered by $\mathcal{V}_{0,4}$. This is due to the fact that Fenchel-Nielsen parameters are the coordinates of the Teichm\"uller space. Moduli space is obtained considering the quotient of mapping class group on the Teich\"uller spcae.  Unfortunately, the action of mapping class group on the Fenchel-Nielsen parameters is very complicated. As a result, we are not able to state the explicit definition of $\mathcal{W}^0_{0,4}$ in terms of $(\ell,\tau)$. In the follow up paper \cite{Moosavian:2017sev}, we have addressed this issue by introducing the notion of an effective description of string vertices.\par

\noindent{\underline{\bf Naive string vertex $\mathcal{V}^0_{1,1}$}}: The naive string vertex $\mathcal{V}^0_{1,1}$ is a collection of once punctured tori with hyperbolic metric on it. The Fuchsian group $\Gamma_{1,1}(\ell,\tau)$ associated with a once punctured torus having Fenchel-Nielsen coordinate $(\ell,\tau)$ can be generated  by the following three elements
\begin{align}\label{Fuchonetorus}
a_1&= \left(\begin{array}{cc}2\text{cosh}\ell &1\\-1 & 0\end{array}\right),\nonumber\\
a_2&= \left(\begin{array}{cc}0 & 1 \\-1 & 2\text{cosh}\ell\end{array}\right),\nonumber\\
a_3&= \frac{1}{\text{sinh}\ell}\left(\begin{array}{cc}e^{\tau}\text{cosh}\ell & e^{\tau} \\e^{-\tau} & e^{-\tau}\text{cosh}\ell\end{array}\right).
\end{align}
The puncture on the torus corresponds to the parabolic element
\begin{align}\label{Fuchelementsotorus}
a_4&= \left(\begin{array}{cc}1 & -4\text{cosh}\ell \\0 & 1\end{array}\right),
\end{align}
whose fixed point is at $\infty$. The local around this fixed point is given by 
\begin{align}\label{v11localc}
w=e^{\frac{\pi^2}{c_*}}e^{2\pi\text{i}z}.
\end{align}
By varying $\ell$ and $\tau$ we can obtain all the string diagrams in $\mathcal{V}^0_{1,1}$. As in the case of $\mathcal{V}^0_{0,4}$, there is no simple description of the domain parameterized by the coordinates $(\ell,\tau)$ that represents $\mathcal{W}^0_{1,1}$, the connected region in the moduli space $\mathcal{M}_{1,1}$ covered by $\mathcal{V}_{0,4}$.

\section{Inconsistency of the naive  string vertices}\label{consistency}

The string vertices $\mathcal{V}^0_{g,n}$ can provide a consistent only if they satisfy the identity (\ref{bvmastercond}).  As we discussed in the subsection (\ref{celldecomp}) the string vertices satisfying the geometrical equation (\ref{bvmastercond}) only if $\mathcal{V}^0_{g,n}$ can give rise to a cell decomposition of the compactified moduli space $\overline{\mathcal{M}}_{g,n}$. Assume that the family of genus $g$ hyperbolic Riemann surfaces with $n$ punctures is parameterized by the Fenchel-Nielsen coordinates. We can claim that the naive string vertex $\mathcal{V}^0_{g,n}$ together with the Feynman diagrams provide  a cell decomposition of $\overline{\mathcal{M}}_{g,n}$  if  the  geodesic lengths and the local coordinates around the punctures on the surfaces at the boundary of $\mathcal{W}^0_{g,n}$, the naive string vertex region inside the moduli space, match exactly with the  geodesic lengths and the local coordinates around the punctures on the surface obtained by the special plumbing fixture construction
\begin{equation}\label{specialplumbing1}
\widetilde z \widetilde w=e^{i\theta} \qquad 0\leq\theta\leq2\pi.
\end{equation}
Here $\widetilde z$ and $\widetilde w$ denote the local coordinates around the punctures that are being glued. Therefore, we must check that the union of $\mathcal{V}^0_{g,n}$ and the regions of the moduli space obtained by the plumbing fixture of different disc neighbourhoods of punctures on the same or different Riemann surfaces belong to $\mathcal{V}^0_{g_i,n_i},$ with appropriate value for $g_i$ and $n_i$, cover the  entire comapctified  moduli space  $\overline{\mathcal{M}}_{g,n}$ exactly once.  \par

Let us  denote  the space of all inequivalent Riemann surfaces with genus $g$ and $n$ punctures obtained by gluing $I$ pairs of punctures on one or more  Riemann surfaces belong to the string vertices $\mathcal{V}^0_{g_i,n_i},$ having appropriate value for $g_i$ and $n_i$, via the plumbing fixture by $V^0_{g,n,I}$. In order to check the consistency of the description of string vertices, we have to check whether the space $\mathcal{V}^0_{g,n}\bigcup V^0_{g,n,1}\bigcup\cdots\bigcup V^0_{g,n,3g-3+n} $ provides a single cover of the compactified moduli space $\overline{\mathcal{M}}_{g,n}$:
\begin{equation}\label{covermodulcell}
\mathcal{V}^0_{g,n}\bigcup V^0_{g,n,1}\bigcup\cdots\bigcup V^0_{g,n,3g-3+n} \mathop{=}\limits^? \overline{\mathcal{M}}_{g,n} 
\end{equation}

In summary, the consistency  of the proposed string vertices  $\mathcal{V}_{g,n}^0$  can be checked by comparing the hyperbolic length of the closed curves  and the induced local coordinates on the surfaces  belong to the boundary of the string vertices with the hyperbolic length of the closed curves and the local coordinates on the surfaces obtained via the plumbing fixture of the surfaces  belong to the  string vertices. To do this, we should first answer the following questions:
\begin{itemize}
	\item What are the hyperbolic lengths of  non-trivial closed curves on the surface obtained via plumbing fixture of the elementary string vertices?
	\item What are the local coordinates induced around the punctures on the surface obtained via plumbing fixture of the elementary string vertices? 
\end{itemize}
For this, we must analyze the hyperbolic metric on the Riemann surfaces obtained via the plumbing fixture of hyperbolic Riemann surfaces.

\subsection{The plumbing fixture vs the cut and paste construction in hyperbolic geometry}\label{degenfixture}

 \begin{figure}
	\begin{center}
		\usetikzlibrary{backgrounds}
		\begin{tikzpicture}[scale=.75]
		     \draw[ thick] (-21,14) to[curve through={(-20.5,13)..(-18,13.95)..(-15.5,13.5)..(-15,14)..(-15.5,15)..(-18,14.05)..(-20.5,14.5)}] (-21,14);
		       \draw[ thick] (-20.5,13.8) to[curve through={(-19.75,13.65)}] (-19.5,13.8);
		     \draw[ thick] (-20.45,13.75) to[curve through={(-19.95,13.95)}] (-19.55,13.75);
		      \draw[ thick] (-16.5,14.3) to[curve through={(-15.75,14.15)}] (-15.5,14.3);
		     \draw[ thick] (-16.45,14.25) to[curve through={(-15.75,14.45)}] (-15.55,14.25);
		     \draw[line width=.75pt,color=red] (-18,14) ellipse (.025 and .05);
		          \draw[line width=.75pt,color=blue,style=dashed] (-18,14) ellipse (.5 and .5);
		     
		       \draw[ thick] (-12,14) to[curve through={(-10.5,13.55)}] (-9,14);
		         \draw[ thick] (-12,13) to[curve through={(-10.5,13.45)}] (-9,13);
		          \draw[line width=.75pt,color=violet] (-12,13.5) ellipse (.2 and .5);
		             \draw[line width=.75pt,color=violet]  (-9,13.5)ellipse (.2 and .5);
		                  \draw[line width=.75pt,color=red] (-10.5,13.5) ellipse (.025 and .05);
		                     \draw[line width=.75pt,color=blue,style=dashed] (-10.5,13.5)  ellipse (1 and 1);
		     
		         \draw[ thick] (-6,14) to[curve through={(-5.5,12.5)..(-3,13)..(-.5,13.5)..(0,14)..(-.5,15)..(-3,14.25)..(-5.5,14.5)}] (-6,14);
		       \draw[ thick] (-5.675,13.975) to[curve through={(-5.35,13.5)}] (-3.55,13.975);
		     \draw[ thick] (-5.55,13.935) to[curve through={(-4.5,14.4)}] (-3.65,13.935);
		      \draw[ thick] (-1.5,14.3) to[curve through={(-0.75,14.15)}] (-.5,14.3);
		     \draw[ thick] (-1.45,14.25) to[curve through={(-0.75,14.45)}] (-.55,14.25);
		     \draw[line width=.75pt,color=red] (-4.25,14.42) ellipse (.025 and .05);
		      \draw[line width=.75pt,color=blue,style=dashed](-4.25,14.42) ellipse (.5 and .5);

				\end{tikzpicture}
	\end{center}
	
	\caption{The two ways of pinching a surface of genus 2; the local model around the pinch is the same, i.e.
a hyperboloid with thin waist.}
	\label{degenerate surface}
\end{figure}

A degenerate Riemann surface is obtained by pinching a non-trivial simple closed curve on the surface. There are two ways of pinching a surface of genus $g$. One way is to pinch a curve along which if we cut, we get two separate Riemann surfaces. Such a degeneration is called a {\it separating degeneration}. Another way is to pinch a curve along which if we cut, we get  a Riemann surface with lower genus and two more boundaries. Such a degeneration is called a {\it non-separating degeneration}. The local model around the pinch for both type of degenerations is the same, i.e. a hyperboloid with thin waist (see figure \ref{degenerate surface}). This limiting case, where the loop degenerates to a point, can be described in terms of the Deligne-Mumford stable curve compactification  of the moduli space of  Riemann surface \cite{mumford2}.  An alternate description for the degenerating families of the hyperbolic Riemann surfaces can  be obtained using the cut and paste construction in the hyperbolic geometry following Fenchel and Nielsen \cite{Maskit4,FN}. In this subsection, we  discuss the relation between these two approaches.

\noindent{\underline{\bf Plumbing fixture construction}}: The moduli space $\mathcal{M}_{g,n}$ of $n$ punctured genus $g$ Riemann surface has several boundaries. Each of these boundaries contains degenerate Riemann surfaces. Adding these degenerate Riemann surfaces to the moduli space produce the compactified moduli space $\overline{\mathcal{M}}_{g,n}$ of  the   the genus $g$ Riemann surface \cite{mumford2}. By definition, a neighbourhood of a node $p$ of $\mathcal{R}$ is complex isomorphic to either $\{|w^{(1)}|<\epsilon\}$ or
\begin{equation}
U=  
\left\{w^{(1)}w^{(2)}=0|~|w^{(1)}|,|w^{(2)}|<\epsilon\right\}
\end{equation}
where $w^{(1)}$ and $w^{(2)}$ are the local coordinates around the  two sides of the node $p$. We can obtain a family of non-degenerate Riemann surfaces from the degenerate Riemann surface $\mathcal{R}$ by  identify $U$ with the 0-fiber of the following family (see figure \ref{Fig.5})
   \begin{equation}
   \{w^{(1)}w^{(2)}=t|~|w^{(1)}|,|w^{(2)}|<\epsilon,|t|<\epsilon\}
   \end{equation}
   A deformation of $\mathcal{R}\in \overline{\mathcal{M}}_g$ which opens the node is given by varying the parameter $t$ (see figure \ref{plumbing1}). \par
  \begin{figure}
	\begin{center}
		\begin{tikzpicture}[scale=.7]
		\draw[thick,color=red] (3,-2.5) ellipse (2 and 2);
		\draw (3,-5.5) node[above]{$D_2$};
		\draw[thick,color=blue] (3,-2.5) ellipse (1 and 1); 
		\draw[thick,color=black](3,-2.5)--(4,-2.5)(3.85,-2.35)--(4,-2.5)(3.85,-2.65)--(4,-2.5)(3.5,-2.2)node{$|t|$};
		
		\draw[thick,color=blue] (-3,-2.5) ellipse (2 and 2);
		\draw (-3,-5.5) node[above]{$D_1$};
		\draw[thick,color=red] (-3,-2.5) ellipse (1 and 1); 
		\draw[thick](-3,-2.5)--(-2,-2.5)(-2.15,-2.35)--(-2,-2.5)(-2.15,-2.65)--(-2,-2.5)(-2.5,-2.2)node{$|t|$};	
		\end{tikzpicture}
	\end{center}
	\caption{ The two annuli having inner radius $|t|$ and outer radius 1 obtained by removing a disc of radius $|t|$ from $D_1$ and $D_2$ where $t$ is a complex parameter.} \label{Fig.5}
\end{figure}

       Let us discuss a more general construction. Consider an arbitrary Riemann surface $\mathcal{R}_0\equiv \mathcal{R}_{(t_1,\cdots,t_m)=(0,\cdots,0)}$ having $m$ nodes.  We denote the $m$  nodes of the degenerate Riemann surface $\mathcal{R}_0$ by $p_1,\cdots,p_m$ . Assume that for each node $p_i$, there is a pair of  punctures $a_i$ and $b_i$ on $\mathcal{R}_0-\{p_1,\cdots,p_m\}$. Consider the following disjoint neighbourhoods  of the punctures $a_i$ and $b_i$ for $i=1,\cdots, m$ 
   \begin{align}
   U_i^1&=\{|w_i^{(1)}|<1\}\nonumber\\ 
   U_i^2&=\{|w_i^{(2)}|<1\}\qquad i=1,\cdots, m
   \end{align}
  Here $w^{(1)}$ and $w^{(2)}_i$ with $w_i^{(1)}(a_i)=0$ and $w_i^{(2)}(b_i)=0$ are the local coordinate around the two sides of the neighbourhood of the node $p_i$.\par
  \begin{figure}
	\begin{center}
		\usetikzlibrary{backgrounds}
		\begin{tikzpicture}[scale=.65]
		\draw[line width=1pt] (-3,1) .. controls (-2.25,1) and (-1.75,.75)  ..(-1,0);
		\draw[line width=1pt] (-3,-2) .. controls(-2.25,-2) and (-1.75,-1.75)  ..(-1,-1);
		\draw[line width=1pt] (-3,.4) .. controls(-1.75,.5)  and (-1.25,-1.2) ..(-3,-1.4);
		\draw[line width=1pt,color=olive] (-3,.7) ellipse (.115 and .315);
		\draw[line width=1pt,color=olive] (-3,-1.7) ellipse (.115 and .315);
		\draw[line width=.75pt,color=blue] (0,-0.5) ellipse (.115 and .155);
		\draw[line width=1pt,color=red] (0.35,-0.5) ellipse (.085 and .115);
		\draw[line width=1pt] (-1,0) .. controls(-.7,-.25) ..(1.5,-0.5)node[above] {$p$};
		\draw[line width=1pt] (-1,-1) .. controls(-.75,-.75)  ..(1.5,-0.5);
		\draw[line width=1pt] (4,0) .. controls(5,1) and (6,1) ..(7,0);
		\draw[line width=1pt] (4,-1) .. controls(5,-2) and (6,-2) ..(7,-1);
		\draw[line width=1pt] (7,0) .. controls(7.2,-.15)  ..(8,-.15);
		\draw[line width=1pt] (7,-1) .. controls(7.2,-.85)  ..(8,-.85);
		\draw[line width=1pt,color=olive] (8,-.5) ellipse (.15 and .35);
		\draw[line width=1pt] (4.75,-.75) .. controls(5.25,-1.15) and (5.75,-1.15) ..(6.25,-.75);
		\draw[line width=1pt] (4.75,-.35) .. controls(5.25,0.1) and (5.75,0.1) ..(6.25,-.35);
		\draw[line width=1pt] (4.75,-.35) .. controls(4.65,-.475) and (4.65,-.625) ..(4.75,-.75);
		\draw[line width=1pt] (6.25,-.35) .. controls(6.35,-.475) and (6.35,-.625) ..(6.25,-.75);
		\draw[line width=1pt] (4,0) .. controls(3.5,-.4) ..(1.5,-0.5)node[below] {$q$};
		\draw[line width=1pt] (4,-1) .. controls(3.5,-.65)  ..(1.5,-0.5);
		\draw[line width=.75pt,color=red] (3.35,-0.525) ellipse (.1 and .145);
		\draw[line width=1pt,color=blue] (3,-0.525) ellipse (.075 and .115);
	
		\draw[->,line width =1pt] (9.5,-.5)--(10.5,-.5);
		\draw[line width=1pt] (12.65,1) .. controls (14.15,.8) and (13.85,.35)  ..(14.65,0);
		\draw[line width=1pt] (12.65,-2) .. controls(14.5,-1.8) and (13.85,-1.35)  ..(14.65,-1);
		\draw[line width=1pt] (12.65,.4) .. controls(13.85,.5)  and (14.40,-1.2) ..(12.65,-1.4);
		\draw[line width=1pt,color=olive] (12.65,.7) ellipse (.115 and .315);
		\draw[line width=1pt,color=olive] (12.65,-1.7) ellipse (.115 and .315);
		\draw[line width=.75pt,color=blue] (15.65,-0.52) ellipse (.115 and .2);
		\draw[line width=1pt] (14.65,0) .. controls(14.9,-.15) ..(16.05,-0.4);
		\draw[line width=1pt] (14.65,-1) .. controls(14.9,-.85)  ..(16.05,-0.65);
		\draw[line width=1pt] (17,0) .. controls(18,1) and (19,1) ..(20,0);
		\draw[line width=1pt] (17,-1) .. controls(18,-2) and (19,-2) ..(20,-1);
		\draw[line width=1pt] (20,0) .. controls(20.2,-.15)  ..(21,-.15);
		\draw[line width=1pt] (20,-1) .. controls(20.2,-.85)  ..(21,-.85);
		\draw[line width=1pt,color=olive] (21,-.5) ellipse (.15 and .35);
		\draw[line width=1pt] (17.75,-.75) .. controls(18.25,-1.15) and (18.75,-1.15) ..(19.25,-.75);
		\draw[line width=1pt] (17.75,-.35) .. controls(18.25,0.1) and (18.75,0.1) ..(19.25,-.35);
		\draw[line width=1pt] (17.75,-.35) .. controls(17.65,-.475) and (17.65,-.625) ..(17.75,-.75);
		\draw[line width=1pt] (19.25,-.35) .. controls(19.35,-.475) and (19.35,-.625) ..(19.25,-.75);
		\draw[line width=1pt] (17,0) .. controls(16.5,-.4) ..(16.05,-0.4);
		\draw[line width=1pt] (17,-1) .. controls(16.5,-.65)  ..(16.05,-0.65);
		\draw[line width=.75pt,color=red] (16.35,-0.54) ellipse (.1 and .15);

		\end{tikzpicture}
	\end{center}
	
	\caption{The plumbing fixture applied on a degenerate Riemann surface with borders around the node represented by the punctures $p$ on the left component surface and $q$ on the right component surface produces a non-degenerate Riemann surface with borders.}
	\label{plumbing1}
\end{figure}

 Then, we  parametrize the opening of the nodes as follows. Given the $m$-tuple
   \begin{equation}
   t=(t_1,\cdots,t_m)\in \mathbb{C}^m,~|t_i|<1 \nonumber
   \end{equation}
    we construct the non-degenerate  Riemann surface $\mathcal{R}_{t}$ as follows. Remove the discs $\{0<|w_i^{(1)}|\leq |t_i|\}$ around the puncture $a_i$ and $\{0<|w^{(2)}_i|\leq |t_i|\}$ around the puncture $b_i$ from the Riemann surface $\mathcal{R}_s$ (see figure \ref{Fig.5}). Then, attach the annular region $\{|t_i|<|w^{(1)}_i|< 1\}$ to the annular region  $\{|t_i|<|w^{(2)}_i|< 1\}$ by identifying $w^{(1)}_i$ and $\frac{t_i}{w^{(2)}_i}$.\par
    
     This construction is complex:  {\it $ t=(t_1,\cdots,t_m)$ parametrizing $\mathcal{R}_{t}$ provides a local complex coordinate chart near the degeneration locus of the compactified moduli space.}\par

         \begin{figure}
\begin{center}
\usetikzlibrary{backgrounds}
\begin{tikzpicture}[scale=1]
\draw[line width=1pt] (1,1) .. controls (1.75,1) and (2.25,.75)  ..(2.75,.2);
\draw[line width=1pt] (1,-2) .. controls(1.75,-2) and (2.25,-1.75)  ..(3,-1);
\draw[line width=1pt] (1,.4) .. controls(1.7,0)  and (1.7,-1) ..(1,-1.4);
\draw[line width=1pt] (1,.7) ellipse (.115 and .315);
\draw[line width=1pt] (1,-1.7) ellipse (.115 and .315);
\draw[line width=1pt] (3,-1) .. controls(3.3,-.75) and (3.75,-.75) ..(4,-1);
\draw[line width=1pt] (4,0.2) .. controls(5,1) and (6,1) ..(7,0);
\draw[line width=1pt] (2.75,0.2) .. controls(3,0.05) and (3.1,0.2) ..(3,1.5);
\draw[line width=1pt] (4,0.2) .. controls(3.8,0.05) and (3.6,0.2) ..(3.7,1.5);
\draw[line width=1pt] (4,-1) .. controls(5,-2) and (6,-2) ..(7,-1);
\draw[line width=1pt] (7,0) .. controls(7.2,-.15)  ..(8,-.15);
\draw[line width=1pt] (7,-1) .. controls(7.2,-.85)  ..(8,-.85);
\draw[line width=1pt] (8,-.5) ellipse (.15 and .35);
\draw[line width=1pt] (3.35,1.5) ellipse (.35 and .15);
\draw[line width=1pt] (4.75,-.75) .. controls(5.25,-1.1) and (5.75,-1.1) ..(6.25,-.75);
\draw[line width=1pt] (4.75,-.35) .. controls(5.25,0.1) and (5.75,0.1) ..(6.25,-.35);
\draw[line width=1pt] (4.75,-.35) .. controls(4.65,-.475) and (4.65,-.625) ..(4.75,-.75);
\draw[line width=1pt] (6.25,-.35) .. controls(6.35,-.475) and (6.35,-.625) ..(6.25,-.75);
\draw[line width =1pt, ]  (2.85,-.48) ellipse (.15 and .65);
\draw[line width =1pt, ] (4.05,-.42) ellipse (.15 and .62);
\draw[line width =1pt] (2.25,0.25) ellipse (.1 and .35);
\draw[line width =1pt] (2.25,-1.25) ellipse (.1 and .35);
\draw[line width =1pt] (1.75,-.55) ellipse (.225 and .1);
\draw[line width =1pt] (5.5,0.375) ellipse (.1 and .365);
\draw[line width =1pt] (5.5,-1.365) ellipse (.1 and .35);
\draw[line width=1pt] (2.2,-.5) ellipse (.25 and .4);

\end{tikzpicture}
\end{center}

\caption{A genus 2 Riemann surface with four borders can be constructed by taking the geometric sum of 6 pairs of pants.}
\label{pairs of pants decomposition}
\end{figure}

\noindent{\underline{\bf Fenchel-Nielsen cut and paste construction}}: The Teichm\"uller  space of the hyperbolic Riemann surfaces  can be parametrized using the Fenchel-Nielsen coordinates \cite{FN}.  The  Fenchel-Nielsen  parametrization is based on the observation that every hyperbolic metric on an arbitrary Riemann surface can be obtained by piecing together the metric from simple subdomains. A compact genus $g$ Riemann surface with $n$ boundary components can be obtained by taking the geometric sum of   $2g-2+n$ pairs of pants (see figure \ref{pairs of pants decomposition}).  The boundary components are the curves with lengths $L_i,~i=1,\cdots,n$.   When all $L_i=0,~i=1,\cdots,n$, we have a genus $g$ Riemann surface with $n$ punctures. \par 

Every  hyperbolic metric on genus $g$ Riemann surface with $n$ borders can be obtained by varying the parameters of this construction. There are two parameters at each attaching site. For the pair of pants $P$ and the pair of pants $Q$, these parameters are the  length $\ell(\beta^P_1)=\ell(\beta^Q_1)\equiv \ell$ of the boundaries $\beta^P_1, \beta^Q_1$ and the twist parameter $\tau$. The twist parameter measures the amount of relative twist performed before glued between the boundaries of the pairs of pants  that are being glued. The precise definition of the twist parameter is  as follows. Let $p_1$  on the boundary $\beta^P_1$ and  $q_1$ on the boundary $\beta^Q_1$ be two points with the following property. The point $p_1$ is the intersection of $\beta_1^P$ and the unique orthogonal geodesic connecting $\beta^P_1$  and  $\beta^P_2$. Similarly, the point $q_1$ is the intersection of $\beta_1^Q$ and the unique orthogonal geodesic connecting $\beta^Q_1$  and $\beta^Q_2$. The twist parameter $\tau$ is the distance between $p_1$ and $q_1$ along $\beta^P_1\sim \beta^Q_1$. Then the parameters 
\begin{equation}
\left(\tau_j,\ell_j\right),~1\leq j\leq 3g-3+n,~\tau_j\in \mathbb{R},~l_j\in \mathbb{R}^+
\end{equation}
 for a fixed pairs of pants decomposition $\mathcal{P}$  endows the Teichm\"uller space $\mathcal{T}_{g,n}$ of the genus-$g$ Riemann surfaces with $n$ boundary components with a global real-analytic coordinates. In this coordinate system,  the Weil-Petersson (WP) symplectic form  takes the following very simple form\cite{Wolpert7}:
\begin{equation}
\omega_{WP}=\sum_{i=1}^{3g-3+n}d\ell_j\wedge d\tau_j \label{WolpertMagic}
\end{equation}

\noindent{\underline{\bf Plumbing fixture vs Fenchel-Nielsen construction}}:  Let us discuss the the relation between the plumbing fixture construction  for the hyperbolic Riemann surfaces  and the cut paste construction of Fenchel and Nielsen when the simple closed geodesic along which we are performing the cut and paste has infinitesimal length. Here, we follow the discussion in \cite{Wolpert1}.  \par

Let us begin by discussing the notion of a collar. For a simple closed geodesic $\alpha$ on the hyperbolic surface $\mathcal{R}$ of length $\ell_{\alpha}$,   the collar around the geodesic $\alpha$ is a neighbourhood around the curve $\alpha$ having area 
\begin{equation}
 2\ell_{\alpha}\text{cot}\frac{\ell_{\alpha}}{2}.
\end{equation}
  The {\it  standard collar} around the geodesic $\alpha$  is the  collection of points $p$ whose hyperbolic distance from the geodesic $\alpha$ is less than  $w(\alpha)$ given by 
	\begin{equation}\label{collarwidth}
	\mathrm{sinh}~w(\alpha)\cdot\mathrm{sinh}~\frac{\ell_{\alpha}}{2}=1.
	\end{equation}
    
         \begin{figure}
\begin{center}
\usetikzlibrary{backgrounds}
\begin{tikzpicture}[scale=.5]
	\begin{pgfonlayer}{nodelayer}
		\node [style=none] (1) at (5.25, -16.25) {};
		\node [style=none] (2) at (5.25, -21.25) {};
		\node [style=none] (3) at (0.25, -21.25) {};
		\node [style=none] (4) at (10.25, -21.25) {};
		\node [style=none] (5) at (3.25, -21.25) {};
		\node [style=none] (6) at (7.25, -21.25) {};
		\node [style=none] (7) at (9, -21.25) {};
		\node [style=none] (8) at (1.25, -21.25) {};
		\node [style=none] (9) at (3.25, -20.25) {};
		\node [style=none] (10) at (1.5, -19.75) {};
		\node [style=none] (11) at (6, -17.5) {};
		\node [style=none] (12) at (1.75, -19) {};
		\node [style=none] (13) at (5, -17.25) {};
		\node [style=none] (14) at (1.25, -20.5) {};
		\node [style=none] (15) at (6.75, -17.75) {};
		\node [style=none] (16) at (5.5, -19.25) {};
		\node [style=none] (17) at (7.5, -18.25) {};
		\node [style=none] (18) at (6.5, -19.5) {};
		\node [style=none] (19) at (8, -18.75) {};
		\node [style=none] (20) at (7, -20) {};
		\node [style=none] (21) at (8.5, -19.25) {};
		\node [style=none] (22) at (7.25, -20.75) {};
		\node [style=none] (23) at (8.75, -20) {};
		\node [style=none] (24) at (7.75, -21.25) {};
		\node [style=none] (25) at (9, -20.75) {};
		\node [style=none] (26) at (10, -19) {};
		\node [style=none] (27) at (14, -19) {};
		\node [style=none] (28) at (15.25, -17.25) {};
		\node [style=none] (29) at (15.25, -20.5) {};
		\node [style=none] (30) at (23.25, -20.5) {};
		\node [style=none] (31) at (23.25, -17.25) {};
		\node [style=none] (32) at (2.75, -21.25) {};
		\node [style=none] (33) at (3.25, -21) {};
		\node [style=none] (34) at (11.5, -18.25) {$w=z^{\frac{2\pi i}{\ell_{\alpha}}}$};
		\node [style=none] (35) at (4.25, -17.25) {};
		\node [style=none] (36) at (4.75, -19.25) {};
		\node [style=none] (37) at (6.25, -17.25) {};
		\node [style=none] (38) at (5.75, -19.25) {};
	\end{pgfonlayer}
	\begin{pgfonlayer}{edgelayer}
		\draw [thick] (3.center) to (4.center);
		\draw [thick] (1.center) to (2.center);
		\draw [thick, bend left=90, looseness=1.75] (5.center) to (6.center);
		\draw [thick, bend left=90, looseness=1.75] (8.center) to (7.center);
		\draw [thick] (8.center) to (9.center);
		\draw [thick] (10.center) to (11.center);
		\draw [thick] (12.center) to (13.center);
		\draw [thick] (14.center) to (15.center);
		\draw [thick] (16.center) to (17.center);
		\draw [thick] (18.center) to (19.center);
		\draw [thick] (20.center) to (21.center);
		\draw [thick] (22.center) to (23.center);
		\draw [thick] (24.center) to (25.center);
		\draw [thick, ->] (26.center) to (27.center);
		\draw [thick, bend right] (28.center) to (31.center);
		\draw [thick, bend left] (29.center) to (30.center);
		\draw [thick, red,bend right=75, looseness=0.25] (28.center) to (29.center);
		\draw [thick, red,bend left=105, looseness=0.25] (28.center) to (29.center);
		\draw [thick, red,bend left=105, looseness=0.25] (31.center) to (30.center);
		\draw [thick, red, bend right=75, looseness=0.25] (31.center) to (30.center);
		\draw [thick] (32.center) to (33.center);
		\draw [red, thick] (35.center) to (36.center);
		\draw [red, thick] (37.center) to (38.center);
	\end{pgfonlayer}
\end{tikzpicture}
\end{center}

\caption{The deck transformation  $z\to e^{\ell_{\alpha}}z$ generates the cyclic group  $\Gamma_{{\alpha}}$.  A fundamental domain for the action of $\Gamma_{{\alpha}}$ is  given by a strip in $\mathbb{H}$. Quotienting $\mathbb{H}$ with $\Gamma_{{\alpha}}$ identify the two sides of the strip. This gives  a hyperbolic  annulus with a hyperbolic metric induced from $\mathbb{H}$.  The local coordinate on the hyperbolic annulus is given by $w=z^{\frac{2\pi i}{\ell_{\alpha}}}$.}
\label{hyperbolic annulus}
\end{figure}

	 The standard   collar can be  described as a quotient of the upper half-plane $\mathbb{H}$. To describe this quotient space, consider the deck transformation 
	 \begin{equation}
	 z\to e^{\ell_{\alpha}}z.
	 \end{equation}
	  It generates a cyclic subgroup of PSL$(2,\mathbb{R})$. Let us  denote this cyclic subgroup by $\Gamma_{{\alpha}}$.  A fundamental domain for the action of $\Gamma_{{\alpha}}$ is  given by a strip in $\mathbb{H}$. When we quotient $\mathbb{H}$ with $z\to e^{\ell_{\alpha}}z$ relation, we identify the two sides of the strip. This gives  a hyperbolic  annulus with a hyperbolic metric induced from $\mathbb{H}$, see figure \ref{hyperbolic annulus}. The core geodesic of this hyperbolic annulus has hyperbolic length $\ell_{\alpha}$. Then the standard collar can be  described as the  quotient of the following  wedge with the cyclic group  $\Gamma_{\alpha}$:
		\begin{equation}
		\left\{\frac{\ell_{\alpha}}{2}<\mathrm{arg} z<\pi-\frac{\ell_{\alpha}}{2}\right\}.
		\end{equation}
	
Let us derive the metric on this hyperbolic annulus. For this, consider a general hyperbolic transformation $h$ that is conjugate to the transformation  $z\to e^{\ell_{\alpha}}z$, with fixed points at $z=x_1$ and $z=x_2$.  The map 
\begin{equation}
g(z)=\frac{z-x_1}{z-x_2}
\end{equation} 
sends fixed points $(x_1,x_2)$ to $(0,\infty)$. As a result, we get
\begin{equation}
\frac{h(z)-x_1}{h(z)-x_2}=e^{\ell_{\alpha}}\frac{z-x_1}{z-x_2}
\end{equation}
Matrix for the transformation $h$ is given by
\begin{equation}
h=\left(\begin{array}{cc}x_1-e^{\ell_{\alpha}}x_2 & (e^{\ell_{\alpha}}-1)x_1x_2 \\1-e^{\ell_{\alpha}} & e^{\ell_{\alpha}}x_1-x_2\end{array}\right)
\end{equation}
The local coordinate that is invariant under this transformation is given by
\begin{equation}
w_h=\left(\frac{z-x_1}{z-x_2}\right)^{\frac{2\pi\mathbf{i}}{\ell_{\alpha}}}\Rightarrow z=\frac{x_1-x_2w^{\frac{\ell_{\alpha}}{2\pi\text{i}}}}{1-w^{\frac{\ell_{\alpha}}{2\pi\text{i}}}}
\end{equation}
A simple computation gives the following expression for the hyperbolic metric in terms of the local coordinate $w_h$. It is given by 
\begin{align}
\frac{dzd\overline{z}}{(\text{Im}z)^2}=\left(\frac{\text{ln}|w_h|\frac{\ell_{\alpha}}{2\pi}}{\text{sinh}\left(\text{ln}|w_h|\frac{\ell_{\alpha}}{2\pi}\right)}\right)^2
\left(\frac{|dw_h|}{|w_h|\text{ln}|w_h|}\right)^2.
\end{align}

	\begin{figure}
	\begin{center}
		\usetikzlibrary{backgrounds}
		\begin{tikzpicture}[scale=.75]
	\begin{pgfonlayer}{nodelayer}
		\node [style=none] (0) at (-12, 5) {};
		\node [style=none] (1) at (-12, 3) {};
		\node [style=none] (2) at (-8, 4) {};
		\node [style=none] (3) at (-7, 4) {};
		\node [style=none] (4) at (-3, 5) {};
		\node [style=none] (5) at (-3, 3) {};
		\node [style=none] (6) at (-11, 4.5) {};
		\node [style=none] (7) at (-11, 3.5) {};
		\node [style=none] (8) at (-4, 4.5) {};
		\node [style=none] (9) at (-4, 3.5) {};
		\node [style=none] (10) at (-2.5, 4) {};
		\node [style=none] (11) at (-0.5, 4) {};
		\node [style=none] (12) at (0, 5.5) {};
		\node [style=none] (13) at (0, 2.5) {};
		\node [style=none] (14) at (2, 4.5) {};
		\node [style=none] (15) at (2, 3.5) {};
		\node [style=none] (16) at (4, 5.5) {};
		\node [style=none] (17) at (4, 2.5) {};
		\node [style=none] (18) at (2, 5.5) {$\alpha$};
		\node [style=none] (19) at (-12, 4) {$ds_0^2$};
		\node [style=none] (20) at (-3, 4) {$ds_0^2$};
		\node [style=none] (21) at (0.75, 4) {$ds_0^2$};
		\node [style=none] (22) at (3.25, 4) {$ds_0^2$};
	\end{pgfonlayer}
	\begin{pgfonlayer}{edgelayer}
		\draw [thick,bend right=15] (0.center) to (2.center);
		\draw [thick,bend left=15] (1.center) to (2.center);
		\draw [thick,bend right=15] (3.center) to (4.center);
		\draw [thick,bend left=15] (3.center) to (5.center);
		\draw [thick, bend left=75, looseness=0.75] (6.center) to (7.center);
		\draw [thick, bend right=60, looseness=0.75] (8.center) to (9.center);
		\draw [thick, bend left=255, looseness=0.75] (6.center) to (7.center);
		\draw [thick, bend left=105] (8.center) to (9.center);
		\draw [->, thick] (10.center) to (11.center);
		\draw [thick] (12.center) to (14.center);
		\draw [thick](13.center) to (15.center);
		\draw [thick](14.center) to (16.center);
		\draw [thick](15.center) to (17.center);
		\draw [thick, bend right=90, looseness=0.75] (14.center) to (15.center);
		\draw [thick, bend left=90, looseness=0.75] (14.center) to (15.center);
	\end{pgfonlayer}
\end{tikzpicture}
	\end{center}
	
	\caption{Plumbing fixture of a pair of cusps produces a collar with curvature accumulated along a curve.}
	\label{collarpfcusp}
\end{figure}
  
Now consider the collar obtained by the plumbing fixture of two punctures with neighbourhoods having local coordinates $w_1$ and $w_2$ on hyperbolic Riemann surfaces with plumbing parameter $t$.  The gluing  produces a collar with curvature accumulated along the curve $w_1=\sqrt{|t|}$. See figure \ref{collarpfcusp}. Therefore, unlike the Fenchel-Nielsen cut and paste construction, the plumbing fixture construction does not produce hyperbolic Riemann surfaces.

 \subsection{ Naive string vertices and the mismatched tiling of the moduli space }\label{gaps1}
 
The observation made in the previous subsection, implies that the naive string vertices $\mathcal{V}^0_{g,n}$ fails to satisfy the geometric equation (\ref{bvmastercond}). To quantify this failure, we must compute the metrics induced on  the Riemann surfaces obtained by the special plumbing fixture (\ref{specialplumbing})  of the hyperbolic Riemann surfaces  and compare with the hyperbolic metric on them.\par

 Consider a set of hyperbolic Riemann surfaces $\mcal{R}^1,\cdots,\mcal{R}^k$.  From this set of surfaces choose $m$-pairs of punctures $p_i,q_i~ i=1,\cdots,m$.   We denote the local coordinates around the punctures $p_i$ and $q_i$ induced from the hyperbolic metric by $w^{(1)}_i$ and $w^{(2)}_i$ with the property that $w^{(1)}_i(p_i)=0$ and $w^{(2)}_i(q_i)=0$. A family of non-degenerate Riemann surfaces $\mcal{R}_t$ parametrized by $m$-tuple $t\equiv (t_1,\cdots,t_m)$ can be constructed by identifying the neighbourhoods of the pairs of punctures $p_i,q_i~i=1,\cdots,m$ using the plumbing fixture relation
	\begin{equation}\label{pl}
	w^{(1)}_iw^{(2)}_i=t_i\qquad i=1,\cdots,m.
	\end{equation}
	
	The hyperbolic metric on $\widehat{\mathcal{R}}_0$, surface obtained by removing the nodes from $\mcal{R}_0$, has the following local expression around the punctures
 \begin{equation}\label{eq:hyperbolic metric around the punctures12}
ds^2=\left(\frac{|d\zeta|}{|\zeta|\ln|\zeta|}\right)^2,\qquad \zeta=w^{(1)}_i,~\zeta=w^{(2)}_i.
 \end{equation}
The plumbing-fixture gluing identifies the curves $w_i^{(1)}=\sqrt{t}$ and $w_i^{(2)}=\sqrt{t}$ in the neighbourhood of the punctures. An important feature of hyperbolic geometry is that gluing two hyperbolic surfaces produces a hyperbolic surface only if they are glued along the geodesics of the surfaces that are being glued \cite{Yoichi}. Since there are no geodesics in the neighbourhood of a puncture on a hyperbolic surface, the curves identified by the plumbing fixture construction are not geodesics on the surfaces that are being glued. As a result the gluing can not induce hyperbolic metric on the resulting surface $\mcal{R}_t$. In fact, one can check that the metric on $\mcal{R}_t$ has constant curvature $-1$ everywhere except along the $w_i^{(1)}=w_i^{(2)}=\sqrt{t}$. Therefore, the plumbing collar does not have hyperbolic metric on it. \par 
	\begin{figure}
	\begin{center}
		\usetikzlibrary{backgrounds}
		\begin{tikzpicture}[scale=.75]
		\draw[line width=1pt]  (2,2) .. controls (0,1.5)  ..(-2,2);
		\draw[line width=1pt]  (2,0) .. controls (0,.5)  ..(-2,0);
		\draw[line width=1pt] (2,1) ellipse (.3 and 1);
		\draw[line width=1pt] (-2,1) ellipse (.3 and 1);
		\draw[line width=1pt,->](2.5,1)--(3.5,1);
		\draw[line width=1pt]  (8,2) --(4,0);
		\draw[line width=1pt]  (8,0) --(4,2);
		\draw[ line width=1pt] (8,1) ellipse (.3 and 1);
		\draw[ line width=1pt] (4,1) ellipse (.3 and 1);
		\end{tikzpicture}
	\end{center}
	
	\caption{The standard collar or the hyperbolic annulus  converging to a pair of cusps.}
	\label{collartocusp}
\end{figure}

However,  a conformal transformation restricted to the plumbing collar the metric on the plumbing collar can convert it into a hyperbolic collar. The resulting hyperbolic collar can be constructed directly via the plumbing fixture of two discs $D_1=\{|y_1|<1\}$ and $D_2=\{|y_2|<1\}$ and by endowing a hyperbolic metric on it.   The  plumbing fixture locus 
 \begin{equation}\label{plumbingfiber}
 \mathcal{F}=\{y_1y_2=t\Big|~|y_1|,|y_2|,|t|<1\}
 \end{equation}
 is a complex manifold fibered  over the disk $D=\{|t|<1\}$. The  $t\neq 0$ fibers are the annuli $\{|t|<|y_1|<1\}$ with complete hyperbolic metric,
 \begin{equation}\label{hype}
  ds_t^2=\text{sinc}^{-2}\left(\frac{\pi\ln|y_1|}{\ln|t|} \right)ds_0^2,
  \end{equation}
 where sinc is the normalized sinc function, $\text{sinc}(x)=\frac{\text{sin}(\pi x)}{\pi x}$, and $ds_0^2$ is the metric on the $t=0$ fiber.  The $t=0$ fiber is the union of the discs $D_1$ and $D_2$ joined at the origin. To obtain a hyperbolic metric, we need to remove the origin from $D_1$ and $D_2$. Then each of the punctured disk has a complete hyperbolic metric given by
 \begin{equation}\label{hypemetricd}   
 ds^2_0=\left(\frac{|dy_1|}{|y_1|\ln|y_1|}\right)^2\qquad \{0<|y_1|<1\}\cup \{0<|y_2|<1\}.
 \end{equation} 
 
	\begin{figure}
	\begin{center}
		\usetikzlibrary{backgrounds}
		\begin{tikzpicture}[scale=.4]
	\begin{pgfonlayer}{nodelayer}
		\node [style=none] (0) at (-13, 5) {};
		\node [style=none] (1) at (-9.75, 3.75) {};
		\node [style=none] (2) at (-7, 4) {};
		\node [style=none] (3) at (-4, 2.25) {};
		\node [style=none] (4) at (-1, 4) {};
		\node [style=none] (5) at (5, 5) {};
		\node [style=none] (6) at (5, -1) {};
		\node [style=none] (7) at (-7, 0) {};
		\node [style=none] (8) at (-13, -1) {};
		\node [style=none] (9) at (-4, 1.5) {};
		\node [style=none] (10) at (-10, 2) {};
		\node [style=none] (11) at (-6, 2) {};
		\node [style=none] (12) at (-2, 2) {};
		\node [style=none] (13) at (2, 2) {};
		\node [style=none] (14) at (-9.75, 0.25) {};
		\node [style=none] (15) at (1.75, 0.25) {};
		\node [style=none] (16) at (-1, 0) {};
		\node [style=none] (17) at (1.75, 3.75) {};
		\node [style=none] (18) at (-4.75, 2.75) {};
		\node [style=none] (19) at (-3.25, 2.75) {};
		\node [style=none] (20) at (-3.25, 1) {};
		\node [style=none] (21) at (-4.75, 1) {};
		\node [style=none] (22) at (-5.5, 3.25) {};
		\node [style=none] (23) at (-5.5, 0.5) {};
		\node [style=none] (24) at (-2.5, 0.5) {};
		\node [style=none] (25) at (-2.5, 3.25) {};
		\node [style=none] (26) at (-8.5, -1.75) {};
		\node [style=none] (27) at (-4.75, 0.75) {};
		\node [style=none] (28) at (-9, 5.5) {};
		\node [style=none] (29) at (-5.5, 3.5) {};
		\node [style=none] (30) at (9, 5) {};
		\node [style=none] (31) at (12.25, 3.75) {};
		\node [style=none] (32) at (15, 4) {};
		\node [style=none] (33) at (18, 2.25) {};
		\node [style=none] (34) at (21, 4) {};
		\node [style=none] (35) at (27, 5) {};
		\node [style=none] (36) at (27, -1) {};
		\node [style=none] (37) at (15.75, 0.25) {};
		\node [style=none] (38) at (9, -1) {};
		\node [style=none] (39) at (18, 1.5) {};
		\node [style=none] (44) at (12.25, 0.25) {};
		\node [style=none] (45) at (23.75, 0.25) {};
		\node [style=none] (46) at (21, 0) {};
		\node [style=none] (47) at (23.75, 3.75) {};
		\node [style=none] (48) at (17.25, 2.75) {};
		\node [style=none] (49) at (18.75, 2.75) {};
		\node [style=none] (50) at (18.75, 1) {};
		\node [style=none] (51) at (17.25, 1) {};
		\node [style=none] (52) at (16.5, 3.25) {};
		\node [style=none] (53) at (16.5, 0.5) {};
		\node [style=none] (54) at (19.5, 0.5) {};
		\node [style=none] (55) at (19.5, 3.25) {};
		\node [style=none] (56) at (17.75, 5.25) {};
		\node [style=none] (57) at (17.25, 3) {};
		\node [style=none] (58) at (18.25, 5.25) {};
		\node [style=none] (59) at (18.75, 3) {};
		\node [style=none] (60) at (5, 2) {};
		\node [style=none] (61) at (9, 2) {};
		\node [style=none] (62) at (17.25, 2.5) {};
		\node [style=none] (63) at (17.25, 1.25) {};
		\node [style=none] (64) at (18.75, 1.25) {};
		\node [style=none] (65) at (18.75, 2.5) {};
		\node [style=none] (66) at (6.75, 2.75) {$f(w)$};
		\node [style=none] (67) at (-9, 6) {$|w_1|= b$};
		\node [style=none] (68) at (-8.5, -2.25) {$|w_1|=e^{-a}b$};
		\node [style=none] (69) at (-4, 6.75) {};
		\node [style=none] (70) at (-4, 2.75) {};
		\node [style=none] (71) at (-4, 7.95) {$|w_1|=|w_2|= \sqrt{|t|}$};
		\node [style=none] (75) at (0.75, -1.75) {};
		\node [style=none] (76) at (-3.25, 0.75) {};
		\node [style=none] (77) at (0.75, -2.25) {$|w_2|=e^{-a}b$};
		\node [style=none] (78) at (-4, 4.25) {};
		\node [style=none] (79) at (1, 5.5) {};
		\node [style=none] (80) at (-2.5, 3.5) {};
		\node [style=none] (81) at (1, 6) {$|w_2|= b$};
		\node [style=none] (82) at (17.75, 6) {discontinuity~of~metric};
		\node [style=none] (83) at (-9.75, 1.75) {};
		\node [style=none] (84) at (-6.25, 1.75) {};
		\node [style=none] (85) at (-1.75, 1.75) {};
		\node [style=none] (86) at (1.75, 1.75) {};
		\node [style=none] (87) at (12, 2) {};
		\node [style=none] (88) at (16, 2) {};
		\node [style=none] (89) at (12.25, 1.75) {};
		\node [style=none] (90) at (15.75, 1.75) {};
		\node [style=none] (91) at (20, 2) {};
		\node [style=none] (92) at (24, 2) {};
		\node [style=none] (93) at (20.25, 1.75) {};
		\node [style=none] (94) at (23.75, 1.75) {};
	\end{pgfonlayer}
	\begin{pgfonlayer}{edgelayer}
		\draw [thick, in=45, out=-45, looseness=1.50] (0.center) to (8.center);
		\draw [thick, in=-150, out=15] (7.center) to (9.center);
		\draw [thick, in=-15, out=150, looseness=0.75] (3.center) to (2.center);
		\draw [thick, in=15, out=165] (2.center) to (1.center);
		\draw [thick, in=-30, out=-165, looseness=0.75] (1.center) to (0.center);
		\draw [thick, in=-165, out=30, looseness=0.75] (3.center) to (4.center);
		\draw [thick, in=-135, out=135, looseness=1.50] (6.center) to (5.center);
		\draw [thick, in=-90, out=-90, looseness=0.75] (10.center) to (11.center);
		\draw [thick, in=-90, out=-90, looseness=0.75] (12.center) to (13.center);
		\draw [thick, style=dashed, bend right=90, looseness=0.50] (3.center) to (9.center);
		\draw [thick, bend left=75, looseness=0.50] (3.center) to (9.center);
		\draw [thick, in=165, out=30, looseness=0.75] (8.center) to (14.center);
		\draw [thick, in=-165, out=-15] (14.center) to (7.center);
		\draw [thick, in=165, out=-30, looseness=1.25] (9.center) to (16.center);
		\draw [thick, in=-165, out=-15, looseness=0.75] (16.center) to (15.center);
		\draw [thick, in=150, out=15, looseness=0.75] (15.center) to (6.center);
		\draw [thick, in=165, out=15] (4.center) to (17.center);
		\draw [thick, in=-150, out=-15, looseness=0.75] (17.center) to (5.center);
		\draw [thick, bend left=60, looseness=0.25] (18.center) to (21.center);
		\draw [thick, style=dashed, bend right, looseness=0.50] (19.center) to (20.center);
		\draw [thick, style=dashed, bend left=225, looseness=0.50] (18.center) to (21.center);
		\draw [thick, bend left=135, looseness=0.50] (19.center) to (20.center);
		\draw [thick, style=dashed, bend right=75, looseness=0.25] (22.center) to (23.center);
		\draw [thick, bend left=75, looseness=0.25] (22.center) to (23.center);
		\draw [thick, style=dashed, bend right=90, looseness=0.25] (25.center) to (24.center);
		\draw [thick, bend left=105, looseness=0.25] (25.center) to (24.center);
		\draw [thick, ->, in=-90, out=90, looseness=0.75] (26.center) to (27.center);
		\draw [thick, ->, in=90, out=-90] (28.center) to (29.center);
		\draw [thick, in=45, out=-45, looseness=1.50] (30.center) to (38.center);
		\draw [thick, in=15, out=165] (32.center) to (31.center);
		\draw [thick, in=-30, out=-165, looseness=0.75] (31.center) to (30.center);
		\draw [thick, in=-135, out=135, looseness=1.50] (36.center) to (35.center);
		\draw [thick, bend right=90, looseness=0.50] (33.center) to (39.center);
		\draw [thick, bend left=75, looseness=0.50] (33.center) to (39.center);
		\draw [thick, in=165, out=30, looseness=0.75] (38.center) to (44.center);
		\draw [thick, in=-165, out=-15] (44.center) to (37.center);
		\draw [thick, in=-165, out=-15, looseness=0.75] (46.center) to (45.center);
		\draw [thick, in=150, out=15, looseness=0.75] (45.center) to (36.center);
		\draw [thick, in=165, out=15] (34.center) to (47.center);
		\draw [thick, in=-150, out=-15, looseness=0.75] (47.center) to (35.center);
		\draw [thick, bend left=60, looseness=0.25] (48.center) to (51.center);
		\draw [thick, style=dashed, bend right, looseness=0.50] (49.center) to (50.center);
		\draw [thick, style=dashed, bend left=225, looseness=0.50] (48.center) to (51.center);
		\draw [thick, bend left=135, looseness=0.50] (49.center) to (50.center);
		\draw [thick, style=dashed, bend right=75, looseness=0.25] (52.center) to (53.center);
		\draw [thick, bend left=75, looseness=0.25] (52.center) to (53.center);
		\draw [thick, style=dashed, bend right=90, looseness=0.25] (55.center) to (54.center);
		\draw [thick, bend left=105, looseness=0.25] (55.center) to (54.center);
		\draw [thick, ->] (56.center) to (57.center);
		\draw [thick, ->] (58.center) to (59.center);
		\draw [thick, ->] (60.center) to (61.center);
		\draw [thick, bend left=15, looseness=0.75] (32.center) to (52.center);
		\draw [thick] (52.center) to (48.center);
		\draw [thick, bend right=15, looseness=0.75] (37.center) to (51.center);
		\draw [thick, in=165, out=-30] (50.center) to (46.center);
		\draw [thick, in=-165, out=30, looseness=0.75] (49.center) to (34.center);
		\draw [thick, bend right] (62.center) to (65.center);
		\draw [thick, bend left] (63.center) to (64.center);
		\draw [thick, bend right=15, looseness=0.50] (62.center) to (63.center);
		\draw [thick, bend left, looseness=0.50] (65.center) to (64.center);
		\draw [thick, ->, in=90, out=-90] (69.center) to (70.center);
		\draw [thick, ->, in=-90, out=90, looseness=0.75] (75.center) to (76.center);
		\draw [thick, ->, in=90, out=-90] (79.center) to (80.center);
		\draw [thick, bend left=90] (83.center) to (84.center);
		\draw [thick, bend left=90] (85.center) to (86.center);
		\draw [thick, in=-90, out=-90, looseness=0.75] (87.center) to (88.center);
		\draw [thick, bend left=90] (89.center) to (90.center);
		\draw [thick, in=-90, out=-90, looseness=0.75] (91.center) to (92.center);
		\draw [thick, bend left=90] (93.center) to (94.center);
	\end{pgfonlayer}
	\end{tikzpicture}
	\end{center}
	
	\caption{ Riemann surface $\mathcal{R}_t$ built by the plumbing fixture of hyperbolic Riemann surfaces has a collar with curvature accumulated along the curve $w_1=w_2=\sqrt{|t|}$. By performing a conformal transformation $f(w)$  which is restricted to a thin collar around the curve $w_1=w_2=\sqrt{|t|}$ on the plumbing collar makes it hyperbolic. On the other hand the metric away from the plumbing collars remain hyperbolic as before the transformation. But these two metrics are do not smoothly join together.  }
	\label{plumbrt}
\end{figure}
The conformal transformation which is restricted to the plumbing collars makes the metric on the plumbing collars on $\mcal{R}_t$ hyperbolic. On the other hand the metric away from the plumbing collars remain hyperbolic as before the transformation. But these two metrics are do not smoothly join together. As a result  the conformal transformation which is restricted to the plumbing collars does not make the metric on $\mcal{R}_t$ hyperbolic, see figure \ref{plumbrt}.  However we can define a smooth metric that matches with the hyperbolic metric on the plumbing collar and the hyperbolic metric on the glued surface away from the plumbing collar by introducing a metric that interpolates between them at the two ends of the plumbing collars.  The resulting metric is a smooth grafted metric $ds_{\text{graft}}^2$ for $\mathcal{R}_t$, see figure \ref{plumbing2}. The grafted metric has curvature $-1$ everywhere except at the tails of the plumbing collar. \par

     \begin{figure}
\begin{center}
\usetikzlibrary{backgrounds}
\begin{tikzpicture}[scale=.4]
	\begin{pgfonlayer}{nodelayer}
		\node [style=none] (0) at (9, 5) {};
		\node [style=none] (1) at (12.25, 3.75) {};
		\node [style=none] (2) at (15, 4) {};
		\node [style=none] (3) at (18, 2.25) {};
		\node [style=none] (4) at (21, 4) {};
		\node [style=none] (5) at (27, 5) {};
		\node [style=none] (6) at (27, -1) {};
		\node [style=none] (7) at (15.75, 0.25) {};
		\node [style=none] (8) at (9, -1) {};
		\node [style=none] (9) at (18, 1.5) {};
		\node [style=none] (10) at (12.25, 0.25) {};
		\node [style=none] (11) at (23.75, 0.25) {};
		\node [style=none] (12) at (21, 0) {};
		\node [style=none] (13) at (23.75, 3.75) {};
		\node [style=none] (14) at (17.25, 2.75) {};
		\node [style=none] (15) at (18.75, 2.75) {};
		\node [style=none] (16) at (18.75, 1) {};
		\node [style=none] (17) at (17.25, 1) {};
		\node [style=none] (18) at (16.5, 3.25) {};
		\node [style=none] (19) at (16.5, 0.5) {};
		\node [style=none] (20) at (19.5, 0.5) {};
		\node [style=none] (21) at (19.5, 3.25) {};
		\node [style=none] (22) at (17.75, 5.25) {};
		\node [style=none] (23) at (17.25, 3) {};
		\node [style=none] (24) at (18.25, 5.25) {};
		\node [style=none] (25) at (18.75, 3) {};
		\node [style=none] (27) at (17.25, 2.5) {};
		\node [style=none] (28) at (17.25, 1.25) {};
		\node [style=none] (29) at (18.75, 1.25) {};
		\node [style=none] (30) at (18.75, 2.5) {};
		\node [style=none] (31) at (17.75, 6) {discontinuity~of~metric};
		\node [style=none] (32) at (12, 2) {};
		\node [style=none] (33) at (16, 2) {};
		\node [style=none] (34) at (12.25, 1.75) {};
		\node [style=none] (35) at (15.75, 1.75) {};
		\node [style=none] (36) at (20, 2) {};
		\node [style=none] (37) at (24, 2) {};
		\node [style=none] (38) at (20.25, 1.75) {};
		\node [style=none] (39) at (23.75, 1.75) {};
		\node [style=none] (40) at (30.5, 5) {};
		\node [style=none] (41) at (33.75, 3.75) {};
		\node [style=none] (42) at (36.5, 4) {};
		\node [style=none] (43) at (39.5, 2.25) {};
		\node [style=none] (44) at (42.5, 4) {};
		\node [style=none] (45) at (48.5, 5) {};
		\node [style=none] (46) at (48.5, -1) {};
		\node [style=none] (47) at (36.5, -0.25) {};
		\node [style=none] (48) at (30.5, -1) {};
		\node [style=none] (49) at (39.5, 1.5) {};
		\node [style=none] (50) at (33.5, 2) {};
		\node [style=none] (51) at (37.5, 2) {};
		\node [style=none] (52) at (41.5, 2) {};
		\node [style=none] (53) at (45.5, 2) {};
		\node [style=none] (54) at (33.75, 0.25) {};
		\node [style=none] (55) at (45.25, 0.25) {};
		\node [style=none] (56) at (42.5, 0) {};
		\node [style=none] (57) at (45.25, 3.75) {};
		\node [style=none] (58) at (38.75, 2.5) {};
		\node [style=none] (59) at (40.25, 2.5) {};
		\node [style=none] (60) at (40.25, 1.25) {};
		\node [style=none] (61) at (38.75, 1.25) {};
		\node [style=none] (62) at (38, 3.25) {};
		\node [style=none] (63) at (38, 0.5) {};
		\node [style=none] (64) at (41, 0.5) {};
		\node [style=none] (65) at (41, 3.25) {};
		\node [style=none] (66) at (35, -1.75) {};
		\node [style=none] (67) at (38.5, 0.75) {};
		\node [style=none] (68) at (34, 5.5) {};
		\node [style=none] (69) at (35.25, 4.5) {};
		\node [style=none] (72) at (27, 2) {};
		\node [style=none] (73) at (31, 2) {};
		\node [style=none] (74) at (29, 2.75) {grafted~metric};
		\node [style=none] (75) at (34, 6) {hyperbolic~metric};
		\node [style=none] (76) at (35, -2.25) {interpolated~metric};
		\node [style=none] (77) at (39.5, 6.75) {};
		\node [style=none] (78) at (39.5, 2.75) {};
		\node [style=none] (79) at (39.5, 7.25) {hyperbolic~annulus};
		\node [style=none] (80) at (44.25, -1.75) {};
		\node [style=none] (81) at (40.5, 0.75) {};
		\node [style=none] (82) at (44.25, -2.25) {interpolated~metric};
		\node [style=none] (83) at (39.5, 4.25) {};
		\node [style=none] (84) at (44, 5.5) {};
		\node [style=none] (85) at (43, 4.5) {};
		\node [style=none] (86) at (44, 6) {hyperbolic~metric};
		\node [style=none] (87) at (33.75, 1.75) {};
		\node [style=none] (88) at (37.25, 1.75) {};
		\node [style=none] (89) at (41.75, 1.75) {};
		\node [style=none] (90) at (45.25, 1.75) {};
		\node [style=none] (91) at (35.5, -0.5) {};
		\node [style=none] (92) at (16.5, -1) {};
		\node [style=none] (93) at (19.5, -1) {};
		\node [style=none] (94) at (18, -1.75) {$\mathcal{F}_b$};
		\node [style=none] (95) at (17.25, 0.5) {};
		\node [style=none] (96) at (18.75, 0.5) {};
		\node [style=none] (97) at (18.25, -.25) {$\mathcal{F}_{e^{-a}b}$};
	\end{pgfonlayer}
	\begin{pgfonlayer}{edgelayer}
		\draw [thick, in=45, out=-45, looseness=1.50] (0.center) to (8.center);
		\draw [thick, in=15, out=165] (2.center) to (1.center);
		\draw [thick, in=-30, out=-165, looseness=0.75] (1.center) to (0.center);
		\draw [thick, in=-135, out=135, looseness=1.50] (6.center) to (5.center);
		\draw [thick, bend right=90, looseness=0.50] (3.center) to (9.center);
		\draw [thick, bend left=75, looseness=0.50] (3.center) to (9.center);
		\draw [thick, in=165, out=30, looseness=0.75] (8.center) to (10.center);
		\draw [thick, in=-165, out=-15] (10.center) to (7.center);
		\draw [thick, in=-165, out=-15, looseness=0.75] (12.center) to (11.center);
		\draw [thick, in=150, out=15, looseness=0.75] (11.center) to (6.center);
		\draw [thick, in=165, out=15] (4.center) to (13.center);
		\draw [thick, in=-150, out=-15, looseness=0.75] (13.center) to (5.center);
		\draw [thick, bend left=60, looseness=0.25] (14.center) to (17.center);
		\draw [thick, style=dashed, bend right, looseness=0.50] (15.center) to (16.center);
		\draw [thick, style=dashed, bend left=225, looseness=0.50] (14.center) to (17.center);
		\draw [thick, bend left=135, looseness=0.50] (15.center) to (16.center);
		\draw [thick, style=dashed, bend right=75, looseness=0.25] (18.center) to (19.center);
		\draw [thick, bend left=75, looseness=0.25] (18.center) to (19.center);
		\draw [thick, style=dashed, bend right=90, looseness=0.25] (21.center) to (20.center);
		\draw [thick, bend left=105, looseness=0.25] (21.center) to (20.center);
		\draw [thick, ->] (22.center) to (23.center);
		\draw [thick, ->] (24.center) to (25.center);
		\draw [thick, bend left=15, looseness=0.75] (2.center) to (18.center);
		\draw [thick] (18.center) to (14.center);
		\draw [thick, bend right=15, looseness=0.75] (7.center) to (17.center);
		\draw [thick, in=165, out=-30] (16.center) to (12.center);
		\draw [thick, in=-165, out=30, looseness=0.75] (15.center) to (4.center);
		\draw [thick, bend right] (27.center) to (30.center);
		\draw [thick, bend left] (28.center) to (29.center);
		\draw [thick, bend right=15, looseness=0.50] (27.center) to (28.center);
		\draw [thick, bend left, looseness=0.50] (30.center) to (29.center);
		\draw [thick, in=-90, out=-90, looseness=0.75] (32.center) to (33.center);
		\draw [thick, bend left=90] (34.center) to (35.center);
		\draw [thick, in=-90, out=-90, looseness=0.75] (36.center) to (37.center);
		\draw [thick, bend left=90] (38.center) to (39.center);
		\draw [thick, in=45, out=-45, looseness=1.50] (40.center) to (48.center);
		\draw [thick, in=15, out=165] (42.center) to (41.center);
		\draw [thick, in=-30, out=-165, looseness=0.75] (41.center) to (40.center);
		\draw [thick, in=-135, out=135, looseness=1.50] (46.center) to (45.center);
		\draw [thick, in=-90, out=-90, looseness=0.75] (50.center) to (51.center);
		\draw [thick, in=-90, out=-90, looseness=0.75] (52.center) to (53.center);
		\draw [thick, style=dashed, bend right=90, looseness=0.50] (43.center) to (49.center);
		\draw [thick, bend left=75, looseness=0.50] (43.center) to (49.center);
		\draw [thick, in=165, out=30, looseness=0.75] (48.center) to (54.center);
		\draw [thick, in=-180, out=-15] (54.center) to (47.center);
		\draw [thick, in=-165, out=-15, looseness=0.75] (56.center) to (55.center);
		\draw [thick, in=150, out=15, looseness=0.75] (55.center) to (46.center);
		\draw [thick, in=165, out=15] (44.center) to (57.center);
		\draw [thick, in=-150, out=-15, looseness=0.75] (57.center) to (45.center);
		\draw [thick, bend left=60, looseness=0.25] (58.center) to (61.center);
		\draw [thick, style=dashed, bend right, looseness=0.50] (59.center) to (60.center);
		\draw [thick, style=dashed, bend left=225, looseness=0.50] (58.center) to (61.center);
		\draw [thick, bend left=135, looseness=0.50] (59.center) to (60.center);
		\draw [thick, style=dashed, bend right=75, looseness=0.25] (62.center) to (63.center);
		\draw [thick, bend left=75, looseness=0.25] (62.center) to (63.center);
		\draw [thick, style=dashed, bend right=90, looseness=0.25] (65.center) to (64.center);
		\draw [thick, bend left=105, looseness=0.25] (65.center) to (64.center);
		\draw [thick, ->, in=-90, out=90, looseness=0.75] (66.center) to (67.center);
		\draw [thick, ->, in=90, out=-90] (68.center) to (69.center);
		\draw [thick, ->] (72.center) to (73.center);
		\draw [thick, ->, in=90, out=-90] (77.center) to (78.center);
		\draw [thick, ->, in=-90, out=90, looseness=0.75] (80.center) to (81.center);
		\draw [thick, ->, in=90, out=-90] (84.center) to (85.center);
		\draw [thick, bend left=90] (87.center) to (88.center);
		\draw [thick, bend left=90] (89.center) to (90.center);
		\draw [thick, in=135, out=-15, looseness=0.75] (42.center) to (62.center);
		\draw [thick, bend right, looseness=0.75] (47.center) to (63.center);
		\draw [thick, bend right=15] (64.center) to (56.center);
		\draw [thick, bend left=15] (65.center) to (44.center);
		\draw [thick, bend right, looseness=1.25] (58.center) to (59.center);
		\draw [thick, bend left] (61.center) to (60.center);
		\draw [thick] (59.center) to (65.center);
		\draw [thick] (60.center) to (64.center);
		\draw [thick] (62.center) to (58.center);
		\draw [thick] (63.center) to (61.center);
		\draw [<->, thick] (92.center) to (93.center);
		\draw [<->, thick] (95.center) to (96.center);
			\end{pgfonlayer}
\end{tikzpicture}

\caption{The smooth grafted metric on $\mathcal{R}_t$ is obtained replacing the metric on the  thin collars at the two edges of the plumbing collar $\mathcal{F}_b$ with an interpolating metric that matches with the hyperbolic metric on the region $\mathcal{F}_{e^{-a}b}$  of plumbing collar and the hyperbolic metric on the glued surface away from the plumbing collar.   The grafted metric has curvature $-1$ everywhere except at the tails of the plumbing collar.}
\label{plumbing2}
\end{center}
\end{figure}
The uniformization theorem asserts that any smooth metric on a Riemann surface can be converted to a hyperbolic metric by applying a suitable conformal transformation. The proper conformal transformation that does this job can be found by solving the so-called {\it the curvature correction equation} \cite{Wolpert2,Wolpert3,Melrose}. To describe this equation, consider a compact Riemann surface with the metric $ds^2$ and the Gauss curvature\footnote{In two dimension, the Gaussian curvature is the half of the Ricci curvature of the surface.} $\mathbf{ C}$.  Then, the  metric $e^{2f}ds^2$ on this surface has constant curvature $-1$ if
  \begin{equation}\label{constantcurvatureq}
  Df-e^{2f}=\mathbf{ C},
  \end{equation}
  where $D$ is the Laplace-Beltrami operator on the surface. \par

In order to find the  hyperbolic metric on $\mathcal{R}_t$, we need a precise definition of the grafted metric. For this, let us introduce arbitrary positive constants $b$ and $a$. The grafted metric $ds_{\text{graft}}^2$ on $\mathcal{R}_t$ is defined as follows:
\begin{itemize}
 \item  On the region complement to the plumbing collars in $\mathcal{R}_t$ described by
    \begin{equation}
   \mathcal{F}_{b}=\left\{\left(w_i^{(1)},w_i^{(2)},t_i\right)\left|~w_i^{(1)}w_i^{(2)}=t_i,|w_i^{(1)}|,|w_i^{(2)}|<b;~i=1,\cdots,m\right\}\right.,
   \end{equation}
   we introduce the hyperbolic metric $ds^2_{\text{hyp}}$ on the surfaces $\mcal{R}^1,\cdots,\mcal{R}^k$ that are being glued.

 \item On the region $ \mathcal{F}_{e^{-a}b}$ in  plumbing collars  $ \mathcal{F}_{b}$  we introduce the hyperbolic metric $ds^2_{t}$  \eqref{hype}.

 \item On the collar bands $\{e^{-a}b\leq |w_i^{(j)}|\leq b\}$ for $j=1,2$, we introduce the following  geometric-interpolation of the hyperbolic metric on $\mcal{R}^1,\cdots,\mcal{R}^k$ and the hyperbolic metric on the plumbing collars     
 \begin{equation}\label{interpmetric}
   ds_{\text{graft}}^2=(ds_{\text{hyp}}^2)^{1-\eta}(ds_t^2)^{\eta} \quad \mathrm{with} \quad \eta=\eta\left(\ln\left(\frac{|w^{(j)}_i|}{b}\right)\right). 
   \end{equation}
   Here $\eta(x)$ is a smooth function that is one for $x\leq -a<0$ and zero for $x\geq 0$.
 
  \end{itemize} 
  
The leading correction to the grafted metric needed for  making it a hyperbolic metric on the plumbing family has already been computed in \cite{Wolpert2,Wolpert3} by solving the curvature correction equation (Theorem $4$ of \cite{Wolpert3}):\\

\noindent{\underline{\bf The expansion of hyperbolic metric on $\mathcal{R}_{t}$}}: {\it Given a choice of $a$ and $b<1$ and a cut-off function $\eta$} {\it, then for all small $t$ the hyperbolic metric $ds^2_{\mathrm{hyp},t}$ on the  Riemann surface $\mathcal{R}_{t}$, obtained by the plumbing fixture of the $m$ pairs of cusps, has the following expansion}:
		\begin{equation}\label{graftedhyp}
		ds^2_{\text{hyp},t}=ds^2_{\text{graft}}\left\{1+\frac{4\pi^4}{3}\sum_{i=1}^m(\ln|t_i|)^{-2}\left(E^{\dagger}_{p_i}+E^{\dagger}_{q_i}\right)+\sum_{i=1}^m\mathcal{O}\left((\ln|t_i|)^{-3}\right)\right\}.
		\end{equation}
		{\it The functions $E^{\dagger}_{p_i}$ and $E^{\dagger}_{q_i}$ are the melding of the Eisenstein series $E(\cdot;2)$ on the component surfaces associated to the pair of cusps (neighbourhoods of punctures $p_i$ and $q_i$) plumbed to form the $i$\textsuperscript{th} collar.}\\
		
We shall describe the function $E^{\dagger}_{p_i}$ associated with the puncture $p_i$ in some detail. Remember that $\mathcal{R}_t$ is obtained by gluing hyperbolic Riemann surfaces $\mcal{R}^1,\cdots,\mcal{R}^j,\cdots,\mcal{R}^k$.  Suppose that the puncture $p_i$ belongs to $\mathcal{R}^j$ whose associated Fuchsian group is  $\Gamma^j$. Also assume that in $\mathbb{H}$ the puncture $p_i$ is represented by the point $\kappa_i$ on the real axis. We denote the stabilizer of $\kappa_i$ in $\Gamma^j$ by $\Gamma^j_{i}$:
 \begin{equation}
 \Gamma^j_{i}=\left\{\sigma\in \Gamma^j\quad |\quad \sigma \kappa_i=\kappa_i\right\}.
 \end{equation}
 Then, the {\it Eisenstein series} $E_{p_i}(z,2)$ defined on $\mathcal{R}^j$ with respect to the puncture $p_i$ is given by 
 \begin{equation}
 E_{p_i}(z,2)=\sum_{\sigma\in \Gamma^j_i\backslash \Gamma^j} \left\{\text{Im}\left(\sigma_i^{-1}\sigma z\right)\right\}^2,
 \end{equation}
  where the transformation $\sigma_i\in \text{SL}(2,\mathbb{R})$  maps $\infty$ to $\kappa_i$: $$\sigma_i\infty =\kappa_i.$$ The transformation $\sigma_i$ is chosen such that $\sigma_i^{-1}\Gamma^j_i\sigma_i$ is equal to the group $\Gamma_{\infty}$ of all matrices of the form    
 $\left(\begin{array}{cc}1 & m \\0 & 1\end{array}\right)$ with $m\in \mathbb{Z}$.  $E_{p_i}(z,2)$ converges locally uniformly on $\mathbb{H}$ and has the expansion
 \begin{equation}\label{eisentseincusp}
 E_{p_i}(z;2)=(\mathrm{Im} ~\sigma_i^{-1}z)^2+\hat e(\sigma_i^{-1}z),
 \end{equation}
 where  $\hat e(\sigma_i^{-1}z)$ is bounded as $\mcal{O}((\mathrm{Im}\, \sigma_i^{-1}z)^{-1})$ for large values of $\mathrm{Im} ~\sigma_i^{-1}z$. The quotient space $\{\mathrm{Im} (\sigma^{-1}_iz)>1\}/\sigma_i^{-1}\Gamma^j_i\sigma_i$ embeds in $\mathbb{H}/\Gamma^j$. This region in $\mathbb{H}$  with hyperbolic area $1$ on $\mathcal{R}^j=\mathbb{H}/\Gamma^j$  is the {\it cusp region} for the neighbourhood of the puncture $p_i$ represented at the infinity of $\mathbb{H}$.  It is useful to consider a special modification of the Eisenstein series for a given choice of $\eta$ and the parameters $b,a$ and $t$. The {\it modified Eisenstein series $E^{\#}_{p_i}$}  is obtained by doing the following modification in the cusp regions
 \begin{itemize}
 \item In the cusp region of the puncture $p_i$ represented at infinity  for $\mathrm{Im} \,\sigma_i^{-1}z>1$, we define
 \begin{equation}\label{trnceisenstein1}
 E_{p_i}^{\#}(z;2)\equiv[1-\eta(-2\pi\mathrm{Im}\,\sigma_i^{-1}z-\mathrm{ln}~b)](\mathrm{Im}\,\sigma_i^{-1}z)^2+\left[1-\eta\left(-2\pi\mathrm{Im}\,\sigma_i^{-1}z+\mathrm{ln}\left(\frac{b}{|t|}\right)-a\right)\right]\hat e(\sigma_i^{-1}z). \nonumber
 \end{equation}
\item In the cusp regions of the other punctures represented at infinity,  for $\mathrm{Im}\, \sigma_i^{-1} z>1$, we define
 \begin{equation}\label{trnceisenstein2}
 E_{p_i}^{\#}(z;2)\equiv \left[1-\eta\left(-2\pi\ln\,\sigma_i^{-1}z+\ln\left(\frac{b}{|t|}\right)-a\right)\right]E(z;2). \nonumber
 \end{equation}
 \end{itemize}
 
Finally we can define the {\it melding of Eisenstein series}  $E_{p_i}^{\dagger}$ on $\mcal{R}_t$. For this, we first extend the definition of $E_{p_i}^{\#}$ by zero on the Riemann surfaces $\mathcal{R}^i,~\forall i\neq j$ that do not contain $p_i$. Then we define $E^{\dagger}$ on the glued surfaces $\mcal{R}_t$ as follows. Away from the plumbing collars in $\mcal{R}_t$, $E^{\dagger}_{p_i}$ is the same as the non-zero $E^{\#}_{p_i}$ in that region. On the $i$\textsuperscript{th} plumbing collar of $\mcal{R}_t$ on the overlap $\{|t|/b<|w^{(1)}|<b\}\cap\{|t|/b<|w^{(2)}|<b\}$, $E^{\dagger}_{p_i}$ is defined as the sum of $E^{\#}_{p_i}$ at $w^{(1)}_i$ and $E^{\#}_{p_i}$ at $w^{(2)}_i=w^{(1)}_i/t$.  \par

  Therefore,  the glued Riemann surface is not a hyperbolic Riemann surface. {\it Only in the  $t\to 0$ limit, we obtain a hyperbolic Riemann surface as a result of the plumbing fixture of hyperbolic Riemann surfaces}.  Hence, the string vertices $ \mathcal{V}^0_{g,n}$ defined as a set of Riemann  surfaces with natural local coordinates induced from the hyperbolic metric  around the punctures do not satisfy the geometrical identity (\ref{bvmastercond}) that is arising from the quantum BV master equation except in the $c_*\to 0$ limit. \par

 Let us elaborate  this. Assume that we obtained a Riemann surface $\widetilde{\mathcal{R}}_{g,n}$ by gluing  two hyperbolic Riemann surfaces  $ \mathcal{R}_{g_i,n_i}$ and  $\mathcal{R}_{g_j,n_j}$ belong to the  string vertices  $ \mathcal{V}^0_{g_i,n_i}$ and $ \mathcal{V}^0_{g_j,n_j}$ respectively via the special plumbing fixture construction (\ref{specialplumbing1}).  The  length $C_*$ of the  geodesic  on the plumbing collar of $\widetilde{\mathcal{R}}_{g,n}$ computed using the hyperbolic metric on the glued surface is given by
\begin{equation}\label{collargeodesica}
C_*=c_*+\mathcal{O}(c_*^{3})
\end{equation}
where $c_*$ is the  length of the geodesic calculated using the grafted metric. Therefore, for the finite values of $c_*$, the geodesics length on the plumbing gets finite corrections. This in particular means that the Fenchel-Nielsen length parameters on the surfaces lying at the boundary of the string vertices and that on the glued surfaces obtained via the special plumbing fixture construction do not match. There is a mismatch of the order  $c_*^3$. \par

We must also  compare the  local coordinates on the surfaces  belong to the boundary of the string vertices with that on the glued surfaces obtained via the special plumbing fixture construction. From  equation (\ref{graftedhyp2}), it is clear that the hyperbolic metric on the surface obtained by gluing $ \mathcal{R}_{g_i,n_i}$ and $ \mathcal{R}_{g_j,n_j}$ do not match with the hyperbolic metric on the relevant regions of $ \mathcal{R}_{g,n}$. Their ratios are different from unity by a term of order $c_*^2$. Therefore the local coordinates on $ \mathcal{R}_{g,n}$ induced from hyperbolic metric deviates from that on the surface obtained by gluing $ \mathcal{R}_{g_i,n_i}$ and $ \mathcal{R}_{g_j,n_j}$ by a term of order $c_*^2$. Thus we  conclude that the naive string vertex $ \mathcal{V}^0_{g,n}$ together with the Feynman diagrams won't be able to provide a single cover  of the moduli space of hyperbolic Riemann surfaces with continuous choice of local coordinates on them. We are left with a mismatch of order $c_*^2$. As a result, {\it the string vertices $\mathcal{V}^0$  provide only a mismatched tiling of the moduli space and the mismatch reduces as we take the parameter $c_*\to 0$}.

\section{Approximately gauge invariant closed string field theory using the corrected string vertices}\label{fillip}

In this section, we  discuss an systematic procedure for improving the approximate cell decomposition of the moduli space  by correcting the definition of the string vertices perturbatively in $c_*$.  We discussed  in the previous section that the reason for the mismatch between the faces of the adjacent cells in the cell decomposition of the moduli space using the string vertices $\mathcal{V}^0$  is that when we glue two hyperbolic surfaces using plumbing fixture, we get a surface which fails to be hyperbolic everywhere. In this section, we argue that the deviation of the induced metric from the hyperbolic metric is of order $c_*^2$. Therefore, the  approximate cell decomposition of the moduli space  can be improved by correcting  the string vertices by modifying the definition of the boundary of the string vertices and the choice of local coordinates around the punctures on the surfaces  belong to the boundary region of the string vertices perturbatively in $c_*$, in a way that  compensate for the deviation  from the hyperbolic metric.

		Using this result, we can obtain the expansion for the hyperbolic metric on $\mathcal{R}_t$ in terms of the grafted metric  (theorem $4$ of \cite{Wolpert3}):
		\begin{equation}
		ds^2_{\text{hyp}}=ds^2_{\text{graft}}\left\{1+\frac{4\pi^4}{3}\sum_{i=1}^m(\ln|t_i|)^{-2}\left(E^{\dagger}_{i,1}+E^{\dagger}_{i,2}\right)+\mathcal{O}\left(\sum_{i=1}^m(\ln|t_i|)^{-3}\right)\right\},
		\label{graftedhyp}
		\end{equation}
		where the functions $E^{\dagger}_{i,1}$ and $E^{\dagger}_{i,2}$ are the melding of the Eisenstein series $E(\cdot;2)$ associated to the pair of cusps plumbed to form the $i$\textsuperscript{th} collar. This expansion for the hyperbolic metric on $\mathcal{R}_{t}$ can be expressed in terms of the length of the $i$\textsuperscript{th} collar geodesic  
		\begin{equation}
		l_{i}=-\frac{2\pi^2}{\ln|t|}+\mathcal{O}\left(\left(\ln|t|\right)^{-2}\right),
		\end{equation}
		 computed using the hyperbolic metric on the annulus as follows:
		\begin{equation}
		ds^2_{\text{hyp}}=ds^2_{\text{graft}}\left(1+\sum_{i=1}^m\frac{l_i^{2}}{3}\left(E^{\dagger}_{i,1}+E^{\dagger}_{i,2}\right)+\mathcal{O}\left(\sum_{i=1}^m\l_i^3\right)\right). 
		\label{graftedhyp2}
		\end{equation}
		Then, the  length of the geodesic in the $i$\textsuperscript{th} plumbing collar  is given by
			\begin{equation}
			l_i^{(\mathrm{hyp})}=-\frac{2\pi}{\ln|t|}\left(1+\mathcal{O}\left(\left(-\ln|t|\right)^{-2}\right)\right)=l_i+\mathcal{O}\left(l_i^3\right),  \label{collargeodesic}
			\end{equation}
			and the  length of a simple closed geodesic $\alpha$,  disjoint from the plumbing collars is given by
			\begin{equation}
			l_{\alpha}\left(\left\{l_{i}\right\}\right)=l_{\alpha}\left(\left\{0\right\}\right)+\sum_{i=1}^m\frac{l_i^2}{6}\int_{\alpha}ds\,\left(E_{i,1}+E_{i,2}\right)+\mathcal{O}\left(\sum_{i=1}^ml_i^3\right).  \label{lengthawayc}
			\end{equation}

			In this formula, $l_{\alpha}\left(\left\{l_{i}\right\}\right)$ is the length of $\alpha$ when the value of the core geodesic of the $i$\textsuperscript{th} collar is $l_i$ computed in the $ds_t^2$ metric which  is given by $l_{i}=-\frac{2\pi^2}{\ln|t|}$ and $l_{\alpha}\left(\{0\}\right)$ means the length of $\alpha$ when the lengths of all plumbing collars are zero.

\subsection{The second order corrections to the string vertices}

Now, we have enough prowess to describe to correct the naive string vertices to second order in $c_*$. Consider the Riemann surface obtained by gluing  $m$ pairs of punctures on a set of hyperbolic Riemann surfaces via the special plumbing fixture construction.  To the second order in $c_*$, we see from the equation (\ref {graftedhyp2}) that the grafted metric $ds^2_{\text{graft}}$ on this  Riemann surface is related to  the hyperbolic metric $ds^2_{\text{hyp}}$ on the Riemann surfaces belong to the boundary of the string vertex $\mathcal{V}_{g,n}^0$  corresponds to $m$ nodes  as follows:
\begin{equation}
ds^2_{\text{hyp}}=ds^2_{\text{graft}}\left(1+\sum_{i=1}^m\frac{c_*^{2}}{3}\left(E^{\dagger}_{i,1}+E^{\dagger}_{i,2}\right)+\mathcal{O}\left(c_*^3\right)\right) \label{graftedhyp2c}
\end{equation}
Using this relation, we  modify the boundary of the  string vertices and the choice of local coordinates on the surfaces belong to a thin neighbourhood of the boundary  of the string vertices as follows.\par

\noindent{\underline{\bf\small  Correction to the boundary of the string vertices}}: The boundary of the string vertex that is obtained by collapsing $m$ propagators  is defined as the locus in the moduli space traced by all the inequivalent hyperbolic Riemann surfaces with $m$ non-homotopic and disjoint  non trivial simple closed curves with length equal to that of the  length of the simple geodesic  on any plumbing collar of a Riemann surface obtained by gluing $m$ pairs of punctures on a set of hyperbolic Riemann surfaces via the special plumbing fixture relation (\ref{specialplumbing1}).  The second order correction to the geodesic length on the plumbing collar of such a Riemann surface can be computed using (\ref{collargeodesic}), and it is given by
	\begin{equation}\label{collargeodesica}
	C_*^{(2)}=c_*+\mathcal{O}\left(c_*^{3}\right)
	\end{equation}
	Therefore, to second order in $c_*$, we use the definition of the region corresponding to the modified string vertex inside the moduli space as the same as that of the naive string vertex, i.e. $\mathcal{W}^2_{g,n}=\mathcal{W}^0_{g,n}$.\par 
	
 \noindent\underline{{\bf\small Correction to the choice of the local coordinates}}: Since there is a modification to the metric, we need to modify the choice of local coordinates around the punctures to make it gluing compatible to second order in $c_*$. For an infinitesimal parameter $\delta$, we modify the local coordinates on the surfaces belong to the naive string vertex as follows.   In order to modify the assignment of  local coordinates in the string vertex $\mathcal{V}^0_{g,n}$, we divide it into subregions. Let us denote the subregion in  $\mathcal{W}^0_{g,n}$ which consists of surfaces with $m$ simple closed geodesics of length between $c_*$ and $(1+\delta)c_*$ by $\mathbf{W}^{(m)}_{g,n}$. Then we modify the local coordinates as follows:
 \begin{itemize}
 \item For surfaces belong to the subregion $\mathbf{W}^{(0)}_{g,n}$, we choose the local coordinate around the $j^{th}$ puncture to be $e^{\frac{\pi^2}{c_*}}w_j$. In terms of $w_j$, the hyperbolic metric in the neighbourhood of the puncture takes the following form
\begin{equation}
 \left(\frac{|dw_j|}{|w_j|\ln|w_j|}\right)^2, \qquad  j=1,\cdots,n. 
\end{equation}
\item For surfaces belong to the region $\mathbf{W}^{(m)}_{g,n}$ with $m\ne 0$, we choose the local coordinates around the $j^{th}$ puncture to be  $e^{\frac{\pi^2}{c_*}}\widetilde{w}_{j,m}$, where $\widetilde{w}_{j,m}$, up to a phase ambiguity,  is given by
\begin{equation}
\widetilde{w}_{j,m}=e^{\frac{c_*^2}{6}\sum_{i=1}^mf(l_i)Y_{ij}}w_{j}.
\end{equation}

\end{itemize}

We found $\widetilde w_{j,m}$ by solving the following equation
\begin{equation}
\left(\frac{|d\widetilde{w}_{j,m}|}{|\widetilde{w}_{j,m}|\mathrm{ln}|\widetilde{w}_{j,m}|}\right)^2=\left(\frac{|dw_j|}{|w_j|\text{ln}|w_j|}\right)^2\left\{1-\frac{c_*^2}{3~\text{ln}|w_j|}\sum_{i=1}^mf(l_i)Y_{ij}\right\},
\end{equation}
where $l_i$ denotes the length of the $i^{th}$ degenerating simple closed geodesic and the function $f(l_i)$ is an arbitrary smooth real function of the geodesic length $l_i$ defined in the interval $\left(c_*,c_*+\delta c_*\right)$, such that $f(c_*)=1$ and $f(c_*+\delta c_*)=0$. The coefficient $Y_{ij}$ is  the leading order term in the following sum around the $j^{th}$ puncture  
\begin{equation}\label{mlocal}
 \text{ln}|w_j|\left(E_{i,1}+E_{i,2}\right)
\end{equation}
where,   $E_{i,1}, E_{i,2}$ denote the Eisenstein series associated with the cusps that are being glued via plumbing fixture to get the collar whose core geodesic is the $i^{th}$ degenerating simple closed geodesic.   The definition and  the expansion of Eisenstein series around a cusp is discussed in appendix (\ref{Eisen}). By using the results discussed there, we obtain $Y_{ij}$ as follows

\begin{align}\label{yij}
Y_{ij}&=\sum_{q=1}^2\sum_{c_i^q,d_i^q}\pi^{2}\frac{\epsilon(j,q)}{|c_i^q|^4}\nonumber\\ c_i^q>0 \qquad &d_i^q~\text{mod}~c_i^q \qquad\left(\begin{array}{cc}* & * \\c_i^q & d_i^q\end{array}\right)\in \quad (\sigma_i^q)^{-1}\Gamma_{i}^{q}\sigma_j
\end{align} 
Here, $\Gamma_i^q$ denotes the Fuchsian group for the component Riemann surface with the cusp  denoted by the index $q$ that is being glued via plumbing fixture to obtain the $i^{th}$ collar.  The transformation $\sigma_j^{-1}$ maps the cusp corresponding to the $j^{th}$ cusp to $\infty$ and $(\sigma_j^q)^{-1}$ maps the cusp denoted by the index $q$ that is being glued via plumbing fixture to obtain the $i^{th}$ collar to $\infty$. The factor $\epsilon(j,q)$ is one if both the $j^{th}$ cusp and he cusp denoted by the index $q$ that is being glued via plumbing fixture to obtain the $i^{th}$ collar belong to the same component surface other wise $\epsilon(j,q)$ is zero.

Let us denote the string vertices corrected in this way by $\mathcal{V}^{2}_{g,n}$. They provide an  improved approximate cell decomposition of the moduli space that has no mismatch up to the order  $c_*^2$. Therefore, to the order  $c_*^2$, the corrected string vertices $\mathcal{V}^{2}_{g,n}$ together with the Feynman diagrams provide an exact cell decomposition of the moduli space. In other words, to second order in $c_*$, the modified string vertices $\mathcal{V}^{2}$ together with the Feynman diagrams defined using the original string vertices $\mathcal{V}^{0}$ provide a single cover of the compactified moduli space. 
\begin{equation}\label{covermodulcellc2}
\mathcal{V}^2_{g,n}\bigcup V^0_{g,n;1}\bigcup\cdots\bigcup V^0_{g,n;3g-3+n} \mathop{=}\limits^{\mathcal{O}(c_*^2)}\overline{\mathcal{M}}_{g,n} 
\end{equation}
	\begin{figure}
	\begin{center}
		\usetikzlibrary{backgrounds}
		\begin{tikzpicture}[scale=.7]
		\draw[line width=1.5pt,->](0,0)--(8,0) node[right]  {$c_*$};
		\draw[line width=1.5pt,->](0,0)--(0,6) node[right]  {$l_{\text{stub}}$};
		\draw[line width=1.5pt,scale=0.55,domain=0.2:9,smooth,variable=\x,red]plot({\x},{-ln(sinh(\x/2))+ln(2)+1/4(1/ sinh(\x/2))(1/ sinh(\x/2))});
		\end{tikzpicture}
	\end{center}
	
	\caption{The length of the stub $l_{\text{stub}}$  increases very fast as the length $c_*$ of the core geodesic on the special plumbing collar becomes small.}
	\label{standardcollarlength}
\end{figure}

Therefore,  the  string vertices $\mathcal{V}^{2}$, corrected perturbatively to the second order in $c_*$, provide a consistent closed string field theory to the order $c_*^2$. In other words,  the corrected string vertices $\mathcal{V}^{2}$ can be used to construct a consistent closed string field theory by keeping $c_*$ very small.  The parameter $c_*$ is related to the length of the stubs used for defining the string vertices. Using the equation for the standard collar width  (\ref{collarwidth}), we can compute the length of the stub. It is given by
\begin{equation}
l_{\text{stub}}=\text{arcsinh}\left(\frac{1}{\text{sinh}\left(c_*/2 \right)} \right)
\end{equation}
From figure (\ref{standardcollarlength}), it is clear that the  length of the stub $l_\text{stub}$  increases very fast as the parameter  $c_*$  becomes small. Therefore, keeping the parameter $c_*$ very small corresponds to adding very long stubs to the string vertices. \par

However, we  emphasize that since the corrected string vertices do not satisfy the geometric realization of the quantum BV master equation beyond order $c_*^2$, the closed string field theory constructed using them will only be approximately gauge invariant.

\subsection{Examples}\label{examples}

We shall explain the procedure for correcting the naive string vertices by correcting the simplest naive string vertices $\mathcal{V}^0_{0,3}$ and $\mathcal{V}^0_{0,4}$ discussed in section (\ref{vertices}). \\

\noindent{\underline{\bf Corrected string vertex $\mathcal{V}^2_{0,3}$}}: The naive string vertex $\mathcal{V}^0_{0,3}$ contain only one surface, which can not be obtained by the plumbing fixture gluing of any other surfaces. Therefore the corrected string vertex $\mathcal{V}^2_{0,3}$ and the naive string vertex $\mathcal{V}^0_{0,3}$ are the same.\\

\noindent{\underline{\bf Corrected string vertex $\mathcal{V}^2_{0,4}$}}:  The naive string vertex $\mathcal{V}^0_{0,4}$ is a collection of four punctured hyperbolic spheres. The surfaces which lie at the boundary of $\mathcal{V}^0_{0,4}$ can be constructed by the special plumbing fixture of two string diagrams that belong to $\mathcal{V}^0_{0,3}$.  In order to make the local coordinates across the boundary of the string vertex continuous we must modify the local coordinates on the string diagrams that form the boundary of the string vertex $\mathcal{V}^0_{0,4}$. For this, let us divide the naive string vertex region $\mathcal{W}^{0}_{0,4}$  into two subregions $\mathbf{W}^{(0)}_{0,4}$ and $\mathbf{W}^{(1)}_{0,4}$. $\mathbf{W}^{(0)}_{0,4}$ region where all the simple closed geodesics on the string diagram has length more than $c_*(1+\delta)$ and $\mathbf{W}^{(1)}_{0,4}=\mathbf{W}^{(1),1}_{0,4}\cup \mathbf{W}^{(1),2}_{0,4}\cup \mathbf{W}^{(0),3}_{0,4}$, where one of the simple closed geodesic on the string diagram has length between $c_*$ and $c_*(1+\delta)$. The simple closed geodesic $\alpha_1$, as shown in  figure \ref{04diagram}, on the string diagrams in $\mathbf{W}^{(1),1}_{0,4}$ that encloses the punctures with marking $1$ and $2$  has length between $c_*$ and $c_*(1+\delta)$. The simple closed geodesic $\alpha_2$ on the string diagrams in $\mathbf{W}^{(1),2}_{0,4}$ that encloses the punctures with marking $1$ and $4$  has length between $c_*$ and $c_*(1+\delta)$. The simple closed geodesic  $\alpha_3$ on the string diagrams in $\mathbf{W}^{(1),3}_{0,4}$ that encloses the punctures with marking $1$ and $3$  has length between $c_*$ and $c_*(1+\delta)$.\par

\begin{figure}
	\begin{center}
		\usetikzlibrary{backgrounds}
		\begin{tikzpicture}[scale=.7]
\begin{pgfonlayer}{nodelayer}
		\node [style=none] (0) at (-2.75, -1.25) {};
		\node [style=none] (1) at (2.5, -1.25) {};
		\node [style=none] (2) at (-0.25, 2.5) {};
		\node [style=none] (3) at (-8, 4.5) {};
		\node [style=none] (4) at (-7.75, 4.75) {};
		\node [style=none] (5) at (7.5, 4.75) {};
		\node [style=none] (6) at (7.75, 4.5) {};
		\node [style=none] (7) at (-0.25, -6.75) {};
		\node [style=none] (8) at (0, -6.75) {};
		\node [style=none] (9) at (-0.25, 1.5) {};
		\node [style=none] (10) at (-1.75, -0.75) {};
		\node [style=none] (11) at (1.5, -0.75) {};
		\node [style=none] (12) at (0, 0) {$\mathbf{W}^{(0)}_{0,4}$};
		\node [style=none] (14) at (2.5, 1.25) {$\mathbf{W}^{(1),1}_{0,4}$};
		\node [style=none] (15) at (-3, 1.25) {$\mathbf{W}^{(1),2}_{0,4}$};
		\node [style=none] (16) at (-0.25, -2.5) {$\mathbf{W}^{(1),3}_{0,4}$};
		\node [style=none] (17) at (1, 0.25) {};
		\node [style=none] (18) at (2, 1) {};
		\node [style=none] (19) at (-1.5, 0.25) {};
		\node [style=none] (20) at (-2.5, 1) {};
		\node [style=none] (21) at (-0.25, -1) {};
		\node [style=none] (22) at (-0.25, -2) {};
		\node [style=none] (23) at (0, 4) {$\overline{\mathcal{M}}_{0,4}$};
		\node [style=none] (24) at (3.75, 3.25) {$1$};
		\node [style=none] (25) at (3.75, 2.25) {$2$};
		\node [style=none] (26) at (4.25, 2.75) {};
		\node [style=none] (27) at (5, 2.75) {};
		\node [style=none] (28) at (5.5, 2.25) {$3$};
		\node [style=none] (29) at (5.5, 3.25) {$4$};
		\node [style=none] (30) at (8.5, 5.5) {s-channel};
		\node [style=none] (31) at (-6, 3.75) {$1$};
		\node [style=none] (32) at (-6, 2.25) {$2$};
		\node [style=none] (33) at (-5.5, 3.25) {};
		\node [style=none] (34) at (-5.5, 3.25) {};
		\node [style=none] (35) at (-5, 2.25) {$3$};
		\node [style=none] (36) at (-5, 3.75) {$4$};
		\node [style=none] (37) at (-5.5, 2.75) {};
		\node [style=none] (38) at (-8.75, 5.5) {u-channel};
		\node [style=none] (39) at (-0.25, -7.5) {t-channel};
		\node [style=none] (40) at (-1, -4.5) {$1$};
		\node [style=none] (41) at (-1, -5.5) {$2$};
		\node [style=none] (42) at (-0.5, -5) {};
		\node [style=none] (43) at (0.25, -5) {};
		\node [style=none] (44) at (0.75, -5.5) {$3$};
		\node [style=none] (45) at (0.75, -4.5) {$4$};
	\end{pgfonlayer}
	\begin{pgfonlayer}{edgelayer}
		\draw [thick](0.center) to (2.center);
		\draw  [thick](2.center) to (1.center);
		\draw  [thick](0.center) to (1.center);
		\draw [thick,bend right] (4.center) to (5.center);
		\draw [thick,bend right=15] (6.center) to (8.center);
		\draw [thick,bend left=15] (3.center) to (7.center);
		\draw  [thick](10.center) to (9.center);
		\draw  [thick](9.center) to (11.center);
		\draw  [thick](11.center) to (10.center);
		\draw  [thick](2.center) to (9.center);
		\draw  [thick](0.center) to (10.center);
		\draw  [thick](11.center) to (1.center);
		\draw [thick, ->] (17.center) to (18.center);
		\draw [thick, ->] (19.center) to (20.center);
		\draw [thick, ->] (21.center) to (22.center);
		\draw  [thick](24.center) to (26.center);
		\draw  [thick](26.center) to (27.center);
		\draw  [thick](27.center) to (29.center);
		\draw  [thick](27.center) to (28.center);
		\draw  [thick](26.center) to (25.center);
		\draw  [thick](31.center) to (33.center);
		\draw  [thick](34.center) to (36.center);
		\draw  [thick](34.center) to (37.center);
		\draw  [thick](37.center) to (32.center);
		\draw  [thick](37.center) to (35.center);
		\draw  [thick](42.center) to (43.center);
		\draw  [thick](43.center) to (44.center);
		\draw  [thick](42.center) to (41.center);
		\draw  [thick](40.center) to (43.center);
		\draw  [thick](42.center) to (45.center);
	\end{pgfonlayer}
\end{tikzpicture}
	\end{center}
	
	\caption{The compactified moduli space of sphere with four punctures is decomposed into the naive string vertex region $\mathcal{W}^{0}_{0,4}$ and the regions filled by the s-channel, u-channel and t-channel string diagrams. The naive string vertex region $\mathcal{W}^{0}_{0,4}$ is further decomposed into $\mathbf{W}^{(0)}_{0,4}$, region where all the simple closed geodesics on the string diagram has length more than $c_*(1+\delta)$ and $\mathbf{W}^{(1)}_{0,4}=\mathbf{W}^{(1),1}_{0,4}\cup \mathbf{W}^{(1),2}_{0,4}\cup \mathbf{W}^{(0),3}_{0,4}$, where one of the simple closed geodesic on the string diagram has length between $c_*$ and $c_*(1+\delta)$. The simple closed geodesic on the string diagrams in $\mathbf{W}^{(1),1}_{0,4}$ that encloses the punctures with marking $1$ and $2$  has length between $c_*$ and $c_*(1+\delta)$. The simple closed geodesic on the string diagrams in $\mathbf{W}^{(1),2}_{0,4}$ that encloses the punctures with marking $1$ and $4$  has length between $c_*$ and $c_*(1+\delta)$. The simple closed geodesic on the string diagrams in $\mathbf{W}^{(1),3}_{0,4}$ that encloses the punctures with marking $1$ and $3$  has length between $c_*$ and $c_*(1+\delta)$.}
	\label{04correctsv}
\end{figure}

\begin{figure}
	\begin{center}
		\usetikzlibrary{backgrounds}
		\begin{tikzpicture}[scale=.6]
	\begin{pgfonlayer}{nodelayer}
		\node [style=none] (0) at (-6, 5) {};
		\node [style=none] (1) at (6, 5) {};
		\node [style=none] (2) at (6, -5) {};
		\node [style=none] (3) at (-6, -5) {};
		\node [style=none] (4) at (-4.75, 3.5) {};
		\node [style=none] (5) at (-4.25, 4) {};
		\node [style=none] (6) at (-4.75, -3.5) {};
		\node [style=none] (7) at (-4, -4) {};
		\node [style=none] (8) at (4.25, 4) {};
		\node [style=none] (9) at (4.5, 3.5) {};
		\node [style=none] (10) at (4.5, -3.5) {};
		\node [style=none] (11) at (4, -4) {};
		\node [style=none] (12) at (0, 2.75) {};
		\node [style=none] (13) at (0, -3.25) {};
		\node [style=none] (14) at (-3.25, 0) {};
		\node [style=none] (15) at (2.75, 0) {};
		\node [style=none] (16) at (-3.75, 2) {};
		\node [style=none] (17) at (-2.5, 3.25) {};
		\node [style=none] (18) at (1.75, -3.25) {};
		\node [style=none] (19) at (3.25, -1.75) {};
		\node [style=none] (20) at (0, 3.5) {$\alpha_1$};
		\node [style=none] (21) at (-4.75, 2.75) {$\alpha_3$};
		\node [style=none] (22) at (3.5, 0) {$\alpha_2$};
		\node [style=none] (23) at (6.75, 5.5) {$4$};
		\node [style=none] (24) at (-6.5, 5.5) {$1$};
		\node [style=none] (25) at (-7, -5.25) {$2$};
		\node [style=none] (26) at (6.75, -5.5) {$3$};
	\end{pgfonlayer}
	\begin{pgfonlayer}{edgelayer}
		\draw [thick, in=210, out=-30, looseness=1.25] (0.center) to (1.center);
		\draw [thick, bend right=45, looseness=1.50] (1.center) to (2.center);
		\draw [thick, bend right=45, looseness=1.25] (3.center) to (0.center);
		\draw [thick, bend left] (3.center) to (2.center);
		\draw [thick, bend left=75] (4.center) to (5.center);
		\draw [style=dashed, thick, bend right=105] (4.center) to (5.center);
		\draw [thick, bend left=90] (8.center) to (9.center);
		\draw [style=dashed, thick, bend right=270, looseness=1.25] (9.center) to (8.center);
		\draw [style=dashed, thick, bend left=75] (6.center) to (7.center);
		\draw [thick, bend right=60] (6.center) to (7.center);
		\draw [style=dashed, thick, bend left=90, looseness=1.25] (11.center) to (10.center);
		\draw [thick, bend right=90, looseness=1.25] (11.center) to (10.center);
		\draw [style=dashed, thick, bend right=90, looseness=0.25] (12.center) to (13.center);
		\draw [thick, bend left=90, looseness=0.25] (12.center) to (13.center);
		\draw [thick, bend right=90, looseness=0.25] (14.center) to (15.center);
		\draw [style=dashed, thick, bend left=90, looseness=0.25] (14.center) to (15.center);
		\draw [thick, bend left=75, looseness=1.50] (16.center) to (17.center);
		\draw [style=dashed, thick, bend right, looseness=0.25] (16.center) to (18.center);
		\draw [style=dashed, thick, bend left=15, looseness=0.75] (17.center) to (19.center);
		\draw [thick, bend right=75, looseness=1.75] (18.center) to (19.center);
		\end{pgfonlayer}
		\end{tikzpicture}
	\end{center}
	
	\caption{By degenerating the simple closed geodesics $\alpha_i,~i=1,2,3$ it is possible to reach the three different boundaries of $\mathcal{M}_{0,4}$.}
	\label{04diagram}
\end{figure}

The punctures on the four punctured sphere  are represented as fixed points of the parabolic elements of the Fuchsian group $\Gamma_{0,4}(\ell,\tau)$ in $\mathbb{H}$. The transformations $a_1,a_2,a_3, a_5,a_6$ and $a_7$ described in (\ref{Gamma04lt}) and (\ref{Gamma04lt1}) are parabolic having fixed points  of them are at $1, -e^{2\tau}, -e^{-\ell+2\tau}, e^{\ell}, e^{-\ell}$ and $-e^{\ell+2\tau}$ respectively. The local coordinate around the puncture can be defined by specifying local coordinates in the neighbourhood of these fixed points. On a string diagram which belongs to  $\mathbf{W}^{(0)}_{0,4}$ around a fixed point at $z=x_j$ we define the local coordinates to be 
\begin{align}\label{v04localc}
w_{x_j}=e^{\frac{\pi^2}{c_*}}e^{-2\pi\text{i}/(z-x_j)}.
\end{align}
On a string diagram which belongs to   $\mathbf{W}^{(1),i}_{0,4}$ around the fixed points $z=x_j$ we choose the following coordinates
\begin{equation}
\widetilde{w}^i_{x_j}=e^{\frac{c_*^2}{6}f(\ell_{\alpha_i})Y_{ji}}w_{x_j}.
\end{equation}
where $\ell_{\alpha_i}$ denotes the length of  $\alpha_i$. The function $f(\ell_{\alpha_i})$ is an arbitrary smooth real function of the geodesic length $\ell_{\alpha_i}$ defined in the interval $\left(c_*,c_*+\delta c_*\right)$, such that $f(c_*)=1$ and $f(c_*+\delta c_*)=0$. The coefficient $Y$ is  given by

\begin{align}\label{yij}
Y_{ji}&=\sum_{c_j^i,d_j^i}\frac{\pi^2}{|c_j^i|^4}\nonumber\\ c_j^i>0 \qquad &d_j^i~\text{mod}~c_j^i \qquad\left(\begin{array}{cc}* & * \\c_j^i & d_j^i\end{array}\right)\in \quad (\sigma_i^q)^{-1}\Gamma(2)\sigma_{x_j}
\end{align} 
 The transformation $\sigma_{x_j}^{-1}$ maps the cusp on the thrice punctured sphere which corresponding to the fixed point $z=x_j$ to $\infty$ and $(\sigma_j^i)^{-1}$ maps the cusp that is being glued via plumbing fixture to obtain the plumbing collar,  on which $\alpha_i$ is the core geodesic, to $\infty$.\par
 
 Due to the complicated action of mapping class group on the Fenchel-Nielsen parameters it is very difficult to find the explicit definition of $\mathbf{W}^{(0)}_{0,4}, \mathbf{W}^{(1),1}_{0,4}, \mathbf{W}^{(0),2}_{0,4}$ and $\mathbf{W}^{(0),3}_{0,4}$ in terms of $(\ell,\tau)$. In the follow up paper \cite{Moosavian:2017sev}, we have addressed this issue by introducing the notion of an effective description of string vertices.

\subsection{Off-shell three point amplitude}\label{off3pamp}

Let us write down the off-shell amplitude associated with the scattering of three off-shell external states represented by the vertex operators $V_1(k_1), V_2(k_2)$ and $V_3(k_3)$ using the string vertices constructed using hyperbolic geometry. It is given by  the following path integral over a thrice punctured sphere with hyperbolic metric on it
\begin{align}\label{3offamp}
A(k_1,k_2,k_3)&= \int \mathcal{D}x^{\mu}\int\mathcal{D}c~\mathcal{D}\overline{c}~\mathcal{D}b~\mathcal{D}\overline{b}~e^{-I_m(x)-I_{gh}(b,c)}\prod_{i=1}^3\left[c\overline{c}~V_{i}(k_{i})\right]_{w_i}\nonumber\\
&=\prod_{i=1}^3\left|\frac{\partial z}{\partial w_i}\right|^{-2}_{w_i=0}\sqrt{\text{det}P_1^{\dagger}P_1}\left(\frac{2\pi^2}{\int d^2z~\sqrt{g}}\text{det}\Delta'\right)^{-13}\int \mathcal{D}x^{\mu}~e^{-I_m(x)}\prod_{i=1}^3\left[V_{i}(k_{i})\right]_{w_i}
\end{align}
where, $\Delta$ is the Laplacian acting on scalars defined on a hyperbolic thrice punctured sphere. The prime indicates that we do not include contributions from zero modes while computing the determinant of $\Delta$. The operator $P_1=\nabla^1_z\oplus\nabla^z_{-1}$ and $P_1^{\dagger}=-\left(\nabla^2_z\oplus\nabla^z_{-2}\right)$. Operators $\nabla^n_z$  and  $\nabla^z_n$ are defined by their action on $T(dz)^n$, which is given by
\begin{align}
\nabla^n_z \left(T(dz)^n\right)&=(g_{z\overline{z}})^n\frac{\partial}{\partial z}\left((g^{z\overline{z}})^nT\right)(dz)^{n+1},\nonumber\\
\nabla^z_n \left(T(dz)^n\right)&=g^{z\overline{z}}\frac{\partial}{\partial\overline{z}}T(dz)^{n-1}.
\end{align}

   Here we assume that  $V_1(k_1), V_2(k_2)$ and $V_3(k_3)$ do not contain any ghost fields. \par

Consider the simplest case where all the three external states are tachyons, and the background spacetime is flat. Then $V_i(k_i)=e^{\text{i} k_i.X},~i=1,2,3$ and $A(k_1,k_2,k_3)$ is given by 
\begin{align}\label{3offampT}
A(k_1,k_2,k_3)&=\prod_{i=1}^3\left|\frac{\partial z}{\partial w_i}\right|^{-2}_{w_i=0}\sqrt{\text{det}P_1^{\dagger}P_1}\left(\frac{2\pi^2}{\int d^2z~\sqrt{g}}\text{det}\Delta'\right)^{-13}\int \mathcal{D}x^{\mu}~e^{-I_m(X)}\prod_{i=1}^3\left[e^{\text{i}k_{i}X_i}\right]_{w_i}\nonumber\\
&=\prod_{i=1}^3\left|\frac{\partial z}{\partial w_i}\right|^{k_i^2-2}_{w_i=0}\sqrt{\text{det}P_1^{\dagger}P_1}\left(\frac{2\pi^2}{\int d^2z~\sqrt{g}}\text{det}\Delta'\right)^{-13}e^{\sum_{i,j}\frac{1}{2}k_{i}\cdot k_jG(x_i,x_j)}\delta(k_1+k_2+k_3),
\end{align}
where $G(x_i,x_j)$ is the Green function for scalars on the hyperbolic thrice punctured sphere. All the quantities appearing in the above expression can be evaluated on any hyperbolic Riemann surface \cite{DHoker:1988pdl}.  The details are discussed in the follow up paper \cite{Moosavian:2017sev}.

\subsection{Off-shell four point tachyon amplitude}\label{off4pamp}
The off-shell four point amplitude has more ingredients compared to the three point amplitude, since it involves an integral over $\overline{\mathcal{M}}_{0,4}$.  For simplicity, let us consider the scattering of four off-shell tachyons. The first step in the calculation is to decompose $\overline{\mathcal{M}}_{0,4}$ following the rules of string field theory:
\begin{equation}\label{Mo4decompose}
\overline{\mathcal{M}}_{0,4}=\mathcal{W}^2_{0,4}\cup F_{0,4}^1\cup F_{0,4}^2\cup F_{0,4}^3,
\end{equation}  
where $\mathcal{W}^2_{0,4}$ is the region in $\overline{\mathcal{M}}_{0,4}$ covered by the modified string vertex ${\mathcal{V}}^2_{0,4}$. $F_{0,4}^1$ is the region covered by the collection of all inequivalent four punctured hyperbolic spheres having a simple closed geodesic $\alpha_1$ which encloses the punctures with marking 1 and 2 and length less than $c_*$, an infinitesimal arbitrary real parameter. $F_{0,4}^2$ is the region covered by the collection of  all inequivalent four punctured hyperbolic spheres having a simple closed geodesic $\alpha_2$ which encloses the punctures with marking 1 and 4 and length less than $c_*$. $F_{0,4}^3$ is region covered by all the collection of four punctured hyperbolic spheres having a simple closed geodesic $\alpha_3$ which encloses the punctures with marking 1 and 3 and length less than $c_*$. The region $\mathcal{W}^2_{0,4}$ has to further divided in order to make a continuous choice of local coordinates on the string diagrams up to order $c_*^2$: $\mathcal{W}^2_{0,4}=\mathbf{W}^{(0)}_{0,4}\cup\mathbf{W}^{(1),1}_{0,4}\cup \mathbf{W}^{(1),2}_{0,4}\cup \mathbf{W}^{(0),3}_{0,4}$. The notations are explained in subsection (\ref{examples}). Then we can write down the off-shell four point tachyon amplitude as follows
\begin{align}\label{4offampT}
&A(k_1,k_2,k_3,k_4)\nonumber\\
&=\int_{\mathbf{W}^{(0)}_{0,4}}d\ell d\tau\prod_{i=1}^4\left|\frac{\partial z}{\partial w_i}\right|^{k_i^2-2}_{w_i=0}\sqrt{\text{det}P_1^{\dagger}P_1}\left(\frac{2\pi^2}{\int d^2z~\sqrt{g}}\text{det}\Delta'\right)^{-13}e^{\sum_{i,j}\frac{1}{2}k_{i}\cdot k_jG(x_i,x_j)}\delta(k_1+k_2+k_3+k_4)\nonumber\\
&+\sum_{r=1}^3\int_{\mathbf{W}^{(1),r}_{0,4}}d\ell d\tau\prod_{i=1}^4\left|\frac{\partial z}{\partial \widetilde{w}^r_i}\right|^{k_i^2-2}_{\widetilde{w}^r_i=0}\sqrt{\text{det}P_1^{\dagger}P_1}\left(\frac{2\pi^2}{\int d^2z~\sqrt{g}}\text{det}\Delta'\right)^{-13}e^{\sum_{i,j}\frac{1}{2}k_{i}\cdot k_jG(x_i,x_j)}\delta(k_1+k_2+k_3+k_4)\nonumber\\
&+\sum_{r=1}^3\int_{F^{r}_{0,4}}d\ell d\tau\prod_{i=1}^4\left|\frac{\partial z}{\partial \widehat{w}^r_i}\right|^{k_i^2-2}_{\widehat{w}^r_i=0}\sqrt{\text{det}P_1^{\dagger}P_1}\left(\frac{2\pi^2}{\int d^2z~\sqrt{g}}\text{det}\Delta'\right)^{-13}e^{\sum_{i,j}\frac{1}{2}k_{i}\cdot k_jG(x_i,x_j)}\delta(k_1+k_2+k_3+k_4),
\end{align}
where $d\ell d\tau$ is Weil-Petersson measure on $\mathcal{M}_{0,4}$ written using the Fenchel-Nielsen coordinates. $G(x_i,x_j)$ is the Green function for scalars on the hyperbolic four punctured sphere. $x_i$ is the fixed point associated with the $i^{\text{th}}$ puncture. These determinants and the Green function can be evaluated on any four punctured hyperbolic sphere \cite{DHoker:1988pdl}. The local coordinates $w_i, \widetilde{w}_i^r$ and $\widehat{w}_i^r$ given by
\begin{align}\label{v04localc1}
w_{i}&=e^{\frac{\pi^2}{c_*}}e^{-2\pi\text{i}/(z-x_i)}\nonumber\\
\widetilde{w}^r_i&=e^{\frac{c_*^2}{6}f(\ell_{\alpha_r})Y_{jr}}w_{i}\nonumber\\
\widehat{w}^r_i&=e^{\frac{\ell_{\alpha_r}^2}{6}Y_{jr}}w_{i}.
\end{align}
The details of $f$ and $Y_{jr}$ are explained in subsection (\ref{examples}). Unfortunately, due to the complicated action of mapping class group on the Fenchel-Nielsen coordinates we don't have explicit description of the integration domains 
$\mathbf{W}^{(0)}_{0,4}, \mathbf{W}^{(1),1}_{0,4}, \mathbf{W}^{(1),2}_{0,4}, \mathbf{W}^{(0),3}_{0,4}, F_{0,4}^1,  F_{0,4}^2$ and $F_{0,4}^3$ in terms of $\ell$ and $\tau$. However, in an interesting way we have resolved this issue in\cite{Moosavian:2017sev}, by introducing the notion of an effective description of string vertices.

\section{Discussions}

In this paper, we constructed the string vertices using Riemann surfaces endowed with metric having constant curvature $-1$ all over the surface. For this we introduced an infinitesimal parameter $c_*$. The parameter $c_*$ is related to the  lengths of the stubs used for defining the string vertices.  The string vertices that we obtained  together with the Feynman diagrams provide a single cover of the moduli space to the order $c_*^2$.  Therefore, by keeping the parameter $c_*$ very small and using the string vertices constructed in this paper, we can obtain a closed string field theory with approximate gauge invariance.\par

Adding stubs to the string vertex refers to the enlargement of the size of the region inside the moduli space that corresponds the string vertex.  Taking $c_*$ very small corresponds to using very long stubs. For constructing a string field theory we are allowed to use stubs having arbitrary length. However, if we choose to add stubs having small length, then we need to find the higher order corrections to the string vertices. We can correct the string vertices up to an arbitrary order by solving the curvature correction equation (\ref{constantcurvatureq}) up to that order. We can then find the corrected string vertex by the procedure introduced in the previous section. Interestingly, the length of stubs  determines  the energy scale of the Wilsonian effective action of the string field theory \cite{Sen:2016qap}.    We would also like to point out that choosing different interpolating functions $f$ and various values for the parameter   $\delta$ give different choices of  local coordinates for elementary string diagrams belongs to the near boundary region of string vertices. However, it is shown in \cite{Pius:2013sca,Pius:2014iaa,Sen:2014pia}  that all such choices of local coordinates give same value for the measurable quantities. 

 In a follow up paper,  we developed these ideas further to provide a calculable framework for the covariant quantum closed bosonic string field theory. In particular, we explained  the rules for the explicit evaluation of the closed bosonic string  field theory  action \cite{Moosavian:2017sev}.  \par
\bigskip
{\bf Acknowledgement:} It is our pleasure to thank Davide Gaiotto and Ashoke Sen for the important comments on the draft and the detailed discussions. We thank   Scott Wolpert  and Barton Zwiebach  for the helpful discussions.  Research at Perimeter Institute is supported by the Government of Canada through Industry Canada and by the Province of Ontario through the Ministry of Research \& Innovation.

\appendix

\section{Brief review of the Batalian-Vilkovisky formalism}\label{BV}

In this section, we  present a brief review of the BV formalism. The construction of an arbitrary gauge theory based on a Lagrangian requires  specifying the basic degrees of freedom  and gauge symmetries. The next step is to construct the action having the specified gauge structure. Finally, quantize the theory by gauge fixing the path integral.  The gauge group of the theory chooses the minimal  procedure that is required for the quantization. For simple gauge groups, like the unitary groups, we can quantize the theory using a relatively simple quantization procedure such as Fadeev-Popov quantization method. However, the gauge group associated with the closed string field theory, namely the homotopy Lie algebra $L_{\infty}$,  endows it with all the features of the most general gauge theory with a Lagrangian description. Therefore, the quantization of such a gauge theory requires the sophisticated machinery of  the BV formalism \cite{Batalin:1981jr,Batalin:1984jr,Barnich:1994db,Barnich:1994mt,Henneaux:1989jq,Henneaux:1992ig,Gomis:1994he}.  \par

The most studied  examples of gauge theories are the non-Abelian Yang-Mills theories with simple gauge groups. The  gauge transformations of such theories  form a simple  Lie groups and  have the following properties:
\begin{itemize}
\item The commutators of the  generators of the Lie group can be expressed as a linear combination of the generators of the Lie group. 
\item The coefficients of the resulting expression, called the structure constants of the algebra, are literally constants. 
\item The algebra of the Lie group is  associative and  satisfies the Jacobi Identities. 
\item All of the above statements are true irrespective of whether the field configuration satisfies the classical equations of motion or not. 
\end{itemize}
A general gauge theory can have more flexible gauge group structure. We are free to allow the following generalizations:
\begin{itemize}
\item The structure constants can be made to depend on the fields involved in the theory with appropriately modified Jacobi Identities.
\item The  gauge transformations itself may have further gauge invariance  that make it a reducible system (see below for the definition of   reducible systems).
\item Two successive gauge transformations can be be allowed to produce another gauge transformation plus a term that vanishes only on-shell.
\end{itemize}

Consider an arbitrary gauge theory with $m_0$ number of gauge invariances whose gauge transformations are not invariant under any other gauge transformation. At the classical level,  we need to introduce a ghost field for each of the $m_0$ gauge invariances.  Assume that the gauge theory also has $m_1$ gauge transformations that keep the $m_0$ gauge transformations  invariant. Suppose that these $m_1$ gauge transformations are not invariant under any further transformations. We call such a gauge theory a first-stage reducible gauge theory. In such theories we need to add $m_1$ ghost for the ghost fields. Therefore, a general $L$\textsuperscript{th}-stage reducible gauge theory with $N$ gauge fields $\phi^i$ has the following set of fields $\Phi^i,~i=1,...,N$  
\begin{equation}\label{setfields}
\Phi^i=\{\phi^i,C^{\alpha_s}_s;~\alpha_s=1,...,m_s ;~s=0,...,L\}
\end{equation}  
where   $C^{\alpha_s}_s$ denotes a ghost field in the theory. With each of these fields let us assign a conserved charge, which we call the ghost number,   as follows. The gauge field $\phi^i$ has the ghost number zero and  the ghost field $C^{\alpha_s}_s$ has the ghost number 
\begin{equation}
gh~[C_s^{\alpha_s}] = s+1
\end{equation}
 Similarly, we can  assign a statistics for each of the ghost fields. The statistics ($\epsilon$) of the ghost field $C_s^{\alpha_s}$ is given by 
\begin{equation}
\epsilon(C^{\alpha_s}_s) = \epsilon_{\alpha_s} +s+1(\mathrm{mod}~2)
\end{equation}
where $ \epsilon_{\alpha_s} $ is the statistics of the level-$s$ gauge parameter. To quantize  a general $L^{\mathrm{th}}$-stage reducible theory, one has to use  the BV quantization procedure. The first step in the BV formalism is the  introduction of a set of  antifields $\Phi^*_i$ for each set of the fields $\Phi^i$. The assignment of the ghost numbers and  the statistics of the antifields are as follows
\begin{equation}\label{antifield}
gh~[\Phi^*_i]=-gh~[\Phi^i]-1,~~~~~ \epsilon(\Phi^*_i)=\epsilon(\Phi^i)+1~(\mathrm{mod}~2)
\end{equation}
Note that a field and its corresponding antifield have  opposite statistics. The  second step is the construction of the classical master action $S[\Phi,\Phi^*]$. The classical master action is a functional of  the fields and the antifields. The  ghost number of the classical action must be  zero and its Grassmanality must be even. The classical master action is required to  satisfy the following  equation known as the {\it classical BV master equation}: 
\begin{equation}\label{masteraction}
\{S,S\}=2\frac{\partial_rS}{\partial\Phi^i}\frac{\partial_lS}{\partial\Phi^*_i}=0
\end{equation}
where $\{,\}$ denotes the antibracket, the subscript $r$ denotes the right derivative and $l$ denotes the left derivative. The left and right derivatives  are defined as  follows
\begin{eqnarray}\label{rleftder}
\frac{\partial_lS}{\partial\Phi^i}&\equiv\frac{\overrightarrow{\partial} S}{\partial\Phi^i} \nonumber
\\
\frac{\partial_rS}{\partial\Phi^i}&\equiv S\frac{\overleftarrow{\partial}}{\partial\Phi^i}
\end{eqnarray}
Assume that $X$ and $Y$ are two functionals of the fields $\Phi^i$ and the antifields $\Phi^*_i$ with the statistics $\epsilon_X$ and $\epsilon_Y$. Then the anti-bracket $\{\cdot,\cdot\}$ is  defined as
\begin{equation}\label{antibracket}
\{X,Y\} \equiv \frac{\partial_rX}{\partial\Phi^i}\frac{\partial_lY}{\partial\Phi^*_i}-\frac{\partial_rX}{\partial\Phi^*_i}\frac{\partial_lY}{\partial\Phi^i}
\end{equation}
The action of the left and the right derivatives on the functional $X$ are related to each other as follows
\begin{equation}
\frac{\partial_lX}{\partial\Phi^i}=(-)^{\epsilon(\Phi^i)(\epsilon_X+1)}\frac{\partial_rX}{\partial\Phi^i}
\end{equation}
However, only those solutions of the classical master equation (\ref{masteraction}) that satisfy  the following set of regularity conditions can be considered as the classical master action $S[\Phi,\Phi^*]$:
\begin{itemize}
\item The classical master action should reduce to the classical action of the gauge theory upon setting all the antifields to zero. This  condition is needed to ensure that we will get back the correct classical limit. 
\item The classical master action should allow the consistent elimination of all the antifields $\Phi^*$. This is needed because antifields are auxiliary fields and they should not be able to make any contribution to the physical observables in the theory.
\end{itemize}
 Such solutions are called {\it the proper solutions} of the classical master equation. It is guaranteed that the classical BV master equation of a general reducible gauge theory has unique proper solutions satisfying these  regularity conditions\cite{Fisch:1989rp}. \par

The meaning of the master equation will be clear once we  expand the BV master action in the antifields. The master equation in the zeroth order in the antifields is the statement of invariance of original action under the gauge transformations. The first order term in the master equation is the algebra satisfied by the gauge transformation. The second order term in the equation is the generalized Jacobi identity and so on. In this sense, BV formalism has the  feature of incorporating the complete structure of the gauge symmetry in the simple looking master equation. \par

The usual BRST formalism allows  the gauge fixed action to have a residual gauge symmetry (the BRST symmetry), whose action is a graded derivation that is nilpotent. Similarly, the  BV formalism also allows the gauge fixed action to have a residual gauge symmetry (the generalized BRST symmetry), whose action is a graded derivation that is nilpotent. The proper solution of the classical BV master equation  has a generalized BRST symmetry even after  gauge-fixing. The generalized BRST transformation, $\delta_B$, of a functional $X$ of fields and antifields generated by a proper solution $S$ is given by
\begin{equation}\label{genbrst}
\delta_BX\equiv \{X,S\}
\end{equation}
The classical master action $S$ is invariant under this transformation due to the classical BV master equation. It is straightforward to check that $\delta_B^2=0$. Therefore, all the classical observables belong to the cohomology of $\delta_B$. \par

Consider the classical master action $S$ of a gauge theory. For any function of $\Upsilon$ of fields, it is straightforward to verify that, the deformed action  
\begin{equation}\label{canonicaltransf}
S'[\Phi,\Phi^*]=S\left[\Phi,\Phi^*+\epsilon \frac{\partial \Upsilon[\Phi]}{\partial \Phi}\right]
\end{equation}
  also satisfies the classical BV-master equation, where  $\epsilon$ is an  arbitrary parameter and $\Upsilon$  is a fermionic functional only of the fields.  Using  this  freedom, we can gauge fix the antifields to $\Phi^*= \frac{\partial \Upsilon[\Phi]}{\partial \Phi}$, and  get rid of the antifields altogether.\par

Finally, we quantize the classical gauge theory by considering the partition function 
\begin{equation}\label{partfunction}
Z_{\Psi} = \int [{\cal D}\Phi]e^{-\frac{1}{\hbar}S\left[\Phi,\frac{\partial \Upsilon[\Phi]}{\partial \Phi}\right]}
\end{equation}
It is important to make sure that physical quantities of the theory do not depend on the choice of the gauge fixing function $\Upsilon$. This is true only if we demand that $S$, the quantum master action, satisfies the {\it the quantum BV-master equation} given by
\begin{align}\label{quantumBV}
\{S,S\} &= -2\hbar \Delta S, ~ \mathrm{at}~ \Phi^*=\frac{\partial \Upsilon[\Phi]}{\partial \Phi}\nonumber\\
\Delta &\equiv \frac{\partial_r}{\partial\Phi^*}\frac{\partial_l}{\partial\Phi}
\end{align}

   \section{Eisenstein Series}\label{Eisen}

 In this appendix, we briefly discuss the definition and some properties of the Eisenstein series following \cite{Kubota}. Consider a discrete subgroup $\Gamma$ of $PSL(2,\mathbb{R})$ acting on the upper half-plane $\mathbb{H}$. Let $\kappa_1,\kappa_2,\cdots,\kappa_h$ be the set of all cusps of $\Gamma$ that are not equivalent with respect to $\Gamma$.  Denote the stabilizer of $\kappa_i$ in $\Gamma$ by $\Gamma_i$:
 \begin{equation}
 \Gamma_i=\left\{\sigma\in \Gamma\quad |\quad \sigma \kappa_i=\kappa_i\right\}
 \end{equation}
 Consider the transformation $\sigma_i\in \text{SL}(2,\mathbb{R})$ which maps $\infty$ to $\kappa_i$: $$\sigma_i\infty =\kappa_i$$ The transformation $\sigma_i$ is chosen such that $\sigma_i^{-1}\Gamma_i\sigma_i$ is equal to the group $\Gamma$ of all matrices of the form    
 $\left(\begin{array}{cc}1 & m \\0 & 1\end{array}\right)$ with $m\in \mathbb{Z}$. Then, the {\it Eisenstein series} $E_i(z,s)$ for the cusp $\kappa_i$ is defined by 
 \begin{equation}
 E_i(z,s)=\sum_{\sigma\in \Gamma_i\backslash \Gamma} \left\{\text{Im}\left(\sigma_i^{-1}\sigma z\right)\right\}^s
 \end{equation}
where $s$ is a complex variable. Whenever the series converges uniformly, the Eisenstein series have the following properties:
\begin{itemize} 
\item $E_i(\sigma z,s)=E_i(z,s)$ for any $\sigma\in \Gamma$;
\item $DE_i=s(s-1)E_i$, where $D$ denotes the Laplacian of $\mathbb{H}$;
\item $E_i$ does not depend on the particular choice of a cusp $x_i$ among equivalent ones;
\item $E_i(z,s)$ converges absolutely, if $\text{Re}(s)>1$.
\end{itemize}

\par

\noindent{\bf{\underline{Fourier expansion at a cusp}~:}} The Fourier expansion of $E_{i}(z,s)$ at $\kappa_j$ is as follows
\begin{align}
E_i(\sigma_j z,s)&=\delta_{ij}\left(\text{Im}~z\right)^s+\phi_{ij}(s)\left(\text{Im}~z\right)^{1-s}\nonumber\\
&+\sum_m 2\pi^s|m|^{s-\frac{1}{2}}\Gamma(s)^{-1} \text{Re}(z)^{\frac{1}{2}}K_{s-\frac{1}{2}}\left(2\pi |m|\text{Re}(z)\right)\phi_{ij,m}(s) e^{2\pi \text{i}m~ \text{Re}(z)}
\end{align}
where $\phi_{ij,m}(s)$ denotes the following summation
\begin{align}
\phi_{ij,m}(s)=\sum_{c,d}\frac{1}{|c|^{2s}}e^{2\pi\text{i}md/c}\qquad\qquad &c>0 \qquad d~\text{mod}~c\nonumber\\& \left(\begin{array}{cc}* & * \\c & d\end{array}\right)\in \quad \sigma_i^{-1}\Gamma\sigma_j
\end{align}
and $\phi_{ij}(s)$ is given by
\begin{equation}
\phi_{ij}(s)=\pi^{\frac{1}{2}}\frac{\Gamma\left(s-\frac{1}{2}\right)}{\Gamma(s)}\phi_{ij,0}(s)
\end{equation}
The matrix $\phi_{ij}(s)$ is symmetric, $\phi_{ij}(s)=\phi_{ji}(s)$.

\end{document}